\documentclass[10pt,journal,compsoc]{IEEEtran}
\IEEEoverridecommandlockouts
\usepackage{hyperref}
\usepackage{amsmath,amssymb,amsfonts}
\usepackage{graphicx,caption}
\usepackage{cleveref}
\usepackage{booktabs}
\usepackage{multirow} 
\usepackage{tabularx}
\usepackage{textcomp}
\usepackage{xcolor}
\usepackage{xspace}
\usepackage[most]{tcolorbox}
\usepackage{booktabs}
\usepackage{listings}
\usepackage{tablefootnote}
\usepackage[nofillcomment,vlined,ruled,linesnumbered]{algorithm2e}
\usepackage{upquote}
\usepackage{amsmath}
\usepackage{subcaption}
\usepackage[T1]{fontenc}
\usepackage[scaled=0.9]{inconsolata} 

\PassOptionsToPackage{svgnames}{xcolor}

\definecolor{codegreen}{rgb}{0,0.6,0}
\definecolor{white}{rgb}{1,1,1}
\definecolor{codegray}{rgb}{0.5,0.5,0.5}
\definecolor{backcolour}{rgb}{0.95,0.95,0.92}
\definecolor{codepurple}{rgb}{0.5,0.0,0.5}

\newcommand{\name}{{\textsc{Libro}}\xspace}
\def\BibTeX{{\rm B\kern-.05em{\sc i\kern-.025em b}\kern-.08em
    T\kern-.1667em\lower.7ex\hbox{E}\kern-.125emX}}

\newcommand{\code}[1]{\lstinline[basicstyle=\normalsize\ttfamily]+#1+}
\newcommand{\codealgo}[1]{{\lstinline[basicstyle=\footnotesize\ttfamily]+#1+}}
\begin{document}

\lstset{
    breaklines=true,
    xleftmargin=1em,                  
    showtabs=false,                  
    tabsize=2,
    numberstyle=\tiny\color{codegray},
    backgroundcolor=\color{white},   
    commentstyle=\color{codegreen},
    keywordstyle=\color{blue},
    stringstyle=\color{codepurple},
    breaklines=true,
    columns=flexible,
    numbers=left,
}

\title{Evaluating Diverse Large Language Models for Automatic and General Bug Reproduction}

\author{Sungmin~Kang*,
        Juyeon~Yoon*,
        Nargiz~Askarbekkyzy,
        and~Shin~Yoo

\IEEEcompsocitemizethanks{
    \IEEEcompsocthanksitem S. Kang, J. Yoon, and S. Yoo are with the Korea Advanced Institute of Science and Technology (KAIST). N. Askarbekkyzy contributed while at KAIST.
    \IEEEcompsocthanksitem S. Kang and J. Yoon contributed equally to this publication.}
}%

\IEEEtitleabstractindextext{%
\begin{abstract}
Bug reproduction is a critical developer activity that is also challenging to automate, 
as bug reports are often in natural language and thus can be difficult to transform to
test cases consistently. As a result, existing techniques mostly focused on crash bugs,
which are easier to automatically detect and verify. In this work, we overcome this
limitation by using large language models (LLMs), which have been demonstrated to be
adept at natural language processing and code generation. By prompting LLMs to generate
bug-reproducing tests, and via a post-processing pipeline to automatically identify
promising generated tests, our proposed technique \name could successfully reproduce
about one-third of all bugs in the widely used Defects4J benchmark. Furthermore, our
extensive evaluation on 15 LLMs, including 11 open-source LLMs, suggests that
open-source LLMs also demonstrate substantial potential, with the StarCoder LLM achieving
70\% of the reproduction performance of the closed-source OpenAI LLM code-davinci-002 on the large Defects4J benchmark,
and 90\% of performance on a held-out bug dataset likely not part of any LLM's training data.
In addition, our experiments on LLMs of different sizes show that bug reproduction using
\name improves as LLM size increases, providing information as to which
LLMs can be used with the \name pipeline.
\end{abstract}

}

\maketitle

\section{Introduction}
\label{sec:intro}

Code will often contain bugs; consequently, bug tracker or bug reporting software
is in popular use among developers~\cite{Anvik2005BugzillaOSS}. When dealing with bug reports submitted
via the tracker, one important activity is bug reproduction~\cite{just2018comparing},
in which a developer tries to reproduce the reported buggy behavior by encapsulating
it in a test case. Such bug-reproducing tests have substantial value both for
developers and automated debugging techniques: prior work has shown that developers
make extensive use of bug reproducing tests (BRTs) when debugging~\cite{Beller2018DebuggingDichotomy}, while many
automated debugging techniques also assume the existence of BRTs as a prerequisite for
operation~\cite{Koyuncu2019ifixr}.

Due to the importance of this task, existing work has suggested automated bug reproduction
techniques, which take a bug report (usually in natural language) as input, and return
a bug-reproducing test as output. However, all such techniques target crash bugs, in which
the test oracle is clear~\cite{Zhao2019na}; in contrast, the research on reproducing functional bugs
has been sparse. This is perhaps due to the difficulty of the level of natural language
processing and code synthesis that is necessary to successfully reproduce a bug: an
automated technique would need to reliably translate the functional specification within
the bug report (which may be in natural language) to an executable test case.

In this work, we first argue that large language models (LLMs), which have demonstrated
substantial performance in natural language processing~\cite{brown2020language} and code synthesis~\cite{Chen2021ec},
may provide a useful solution to this problem. To this end, we propose \name, a pipeline
that uses LLMs to first generate a number of BRT candidates via constructing an appropriate
prompt with a given bug report. All generated tests are subsequently evaluated by the
postprocessing pipeline of \name. It will automatically filter low-quality tests that
do not fail, and sort the remaining tests based on test execution features to minimize
developer inspection cost of automatically generated tests; this is critical for improving
developer experience and adoption~\cite{Ohearn2020ICSEKeynote}.

Our experiments reveal that when using the code-davinci-002 model from OpenAI, 
\name can reproduce one-third of all the bugs in the widely
used Defects4J benchmark~\cite{Just:2014aa}, substantially outperforming bug reproduction baselines
that we compared against. Furthermore, our postprocessing pipeline could successfully
identify cases where \name was more likely to yield accurate bug reproduction results, allowing
\name to control the precision of \name up to 90\% and to place an actual BRT as the
first-place suggestion in 60\% of all bugs that could be reproduced.

This work is an extension of our previous publication~\cite{Kang2023LIBRO}. Relative to
our previous publication, this extension primarily contributes a newly performed extensive
study comparing the capabilities of 15 LLMs in total, including 11 open-source LLMs; to the best of our knowledge, we are the first to perform such a large-scale comparison of open source LLMs on testing.
Our LLM comparison also allows us to perform multiple analyses that were impossible to do in our previous work: 
(1) we plot the trade-off between model GPU memory consumption and performance, which can help practitioners decide which LLM to use given their computational resources; 
(2) evaluate the influence of model size on performance when the training technique is controlled; 
(3) discuss how the output of ChatGPT has changed over time, and 
(4) explore whether self-consistency~\cite{wang2023selfconsistency}, the basis for our post-processing pipeline, works for other LLMs as well.
The LLM comparisons made in this paper required more than eight months of GPU time and seven months of CPU time to be fully executed. The
specifics of our findings are detailed below:

\begin{itemize}
    \item Through experiments with multiple LLMs, we demonstrate that \name is a general technique that is not only effective when using a particular LLM.
    \item Through our comparison of 15 LLMs, we find that among the closed-source LLMs which we could experiment with, the \texttt{code-davinci-002} model showed the best performance, while among the open-source LLMs, the recent StarCoder model~\cite{li2023starcoder} showed the best performance, reproducing about 70\% of the bugs that \texttt{code-davinci-002} could reproduce. Furthermore, we demonstrate that open-source LLMs can make bug-reproducing tests outside of their training data.
    \item By comparing LLMs from the same family trained with different data or parameters, we find that fine-tuning code LLMs on natural language can hurt performance, and that larger models tend to show better performance, suggesting guidelines on what code LLMs to use when operating \name.
    \item We experiment with the temperature parameter of LLMs, which controls the randomness of the LLM output, and find that a value of 0.6 works best.
    \item We make our experimental data and analysis scripts publicly available: \url{https://github.com/coinse/libro-journal-artifact}.
\end{itemize}

The remainder of the paper is organized as follows. We motivate our research in \Cref{sec:motivation}. Based on this, we describe our approach
in \Cref{sec:approach}. Evaluation settings and research questions are in \Cref{sec:evaluation}
and \Cref{sec:rqs}, respectively. Results are presented in \Cref{sec:results}, \Cref{sec:discussion} provides in-depth discussion of our technique, \Cref{sec:relwork} gives an overview of the
relevant literature, and \Cref{sec:conclusion} concludes.

\section{Motivation}
\label{sec:motivation}

\subsection{Bug Reproduction}

As described in the previous section, generating bug-reproducing tests from
bug reports is important for both developers and automated techniques. First, 
Koyuncu et al.~\cite{Koyuncu2019ifixr} report that Spectrum-Based Fault Localization (SBFL) 
techniques cannot locate the bug at the time of being reported in 95\% of the 
cases they analyzed, due to the lack of a bug-reproducing test when the report
was first filed. Other studies show that developers use bug-reproducing tests
extensively when doing debugging themselves: for example, Beller et al.~\cite{Beller2018DebuggingDichotomy}
note that 80\% of developers would use bug-reproducing tests to verify fixes.
Automatic bug reproduction is also important as the report-to-test problem is a perhaps 
underappreciated yet nonetheless important and recurring part of testing. Kochhar et al.~\cite{
Kochhar2019Practitioner}, for example, explicitly ask hundreds of developers on whether they 
agree to the statement ``during maintenance, when a bug is fixed, it is good to 
add a test case that covers it'', and find a strong average agreement of 4.4 on 
a Likert scale of 5.

To further verify that developers regularly deal with the report-to-test 
problem, we analyze the number of test additions that can be attributed to 
a bug report, by mining hundreds of open-source Java repositories. We start 
with the \texttt{Java-med} dataset from Alon et al.~\cite{Alon2019ty},
which consists of 1000 top-starred Java projects from GitHub. From the list of 
commits in each repository, we check (i) whether the commit adds a test, and 
(ii) whether the commit is linked to an issue. To determine whether a commit 
adds a test, we check that its diff adds the \texttt{@Test} decorator along 
with a test body. In addition, we link a commit to a bug report (or an 
\emph{issue} in GitHub) if (i) the commit message mentions 
"(fixes/resolves/closes) \#NUM", or (ii) the commit message mentions a pull 
request, which in turn mentions an issue. We compare the number of tests added 
by such report-related commits to the size of the test 
suite at the time of gathering (August 2022) 
to estimate the prevalence of such tests. As different repositories
have different issue-handling practices, we filter out repositories that have
no issue-related commits that add tests, as this indicates a different
bug handling practice (e.g. \texttt{google/guava}). Accordingly, we analyze
300 repositories, as shown in \Cref{tab:repo_categories}.

\begin{table}[ht]
    \centering
    \caption{Analyzed repository characteristics\label{tab:repo_categories}}
    \scalebox{0.9}{
    \begin{tabular}{lc}
    \toprule
    Repository Characteristic & \# Repositories\\\midrule
    Could be cloned & 970 \\
    Had a JUnit test (\texttt{@Test} is found in repository) & 550 \\
    Had issue-referencing commit that added test & 300 \\
    \bottomrule
    \end{tabular}
    }
\end{table} 

Overall, in these 300 repositories that we inspected, the number of tests that were
added by issue-referencing commits was on median 28.4\% of the overall test suite size. While
one must note that due to code evolution, this may not mean that 28.4\% of all current
tests were added by bug reports, overall these results
suggest a substantial portion of bugs are being added via bug reports, supporting
our claim that test addition after bug reproduction is a common task that developers
face. In turn, automating the report-to-test activity could provide substantial
benefit to developers, and would help them within their existing workflow.

However,
the problem has been difficult for researchers to tackle until recently,
due to the difficulties of natural language processing and corresponding
code synthesis. For example, the bug report in 
\Cref{tab:ex-report-math-63} does not explicitly specify any code, which would make
this report difficult to automatically process for traditional techniques, even though a user
fluent in English and Java would be capable of deducing that when both arguments
are NaN, the `equals' methods in `MathUtils' should return \code{false}. 
As a result, most existing work focused on reproducing
crashes~\cite{Soltani2020aa, Nayrolles2015Jcharming}.

Large Language Models (LLMs), which have been pre-trained on large
corpora of natural language and programming language data, may provide a
potential solution to tame this difficulty. LLMs show a
surprising level of performance on both natural language processing~\cite{brown2020language}
and software engineering~\cite{Kang2023LIBRO,lemieux2023codamosa} tasks.
Their capability, and the recent introduction of many different open-source LLMs
as described in the next subsection, thus poses the question of whether this
emerging technology could be used to alleviate developer effort in writing tests from bug reports.

\subsection{Open-source Large Language Models}
While Brown et al.~\cite{brown2020language} first demonstrated the substantial
potential of LLMs, the LLMs from OpenAI have not since been made public - that is,
they have only been made available via an API, without any sharing of source code
or neural network weights. This
makes consistent research using LLMs difficult, as LLMs accessed through an API
can see significant behavior change in a short period of time, as Chen et al.~\cite{chen2023chatgptchange}
identify. Indeed, earlier this year, the API to the \texttt{code-davinci-002} model was
discontinued~\cite{OpenAICodexDiscontinued}, making previous papers that utilized the model (including
our previous work~\cite{Kang2023LIBRO}) difficult to reproduce. In reaction to the secrecy
and potential centralization threatening the open spirit of research, multiple
open-source LLMs have been suggested~\cite{workshop2023bloom,li2023starcoder}. Unlike the widely-used OpenAI models, which require a per-usage
fee for their API use and which are not available in all countries, the LLMs that
we use are generally free for use in research in all countries, and can be operated
without cost provided that one has the necessary computational resources. As a
result, while much research is using OpenAI models such as ChatGPT~\cite{OpenAICodexDiscontinued}, we believe that
it would benefit the software engineering community to evaluate open-source LLMs under the same
task as closed-source LLMs and compare their performance, to showcase their viability
and promote their widespread use in research.

\section{Approach}
\label{sec:approach}

\begin{figure}[h!]
\centering
\includegraphics[width=1.0\linewidth]{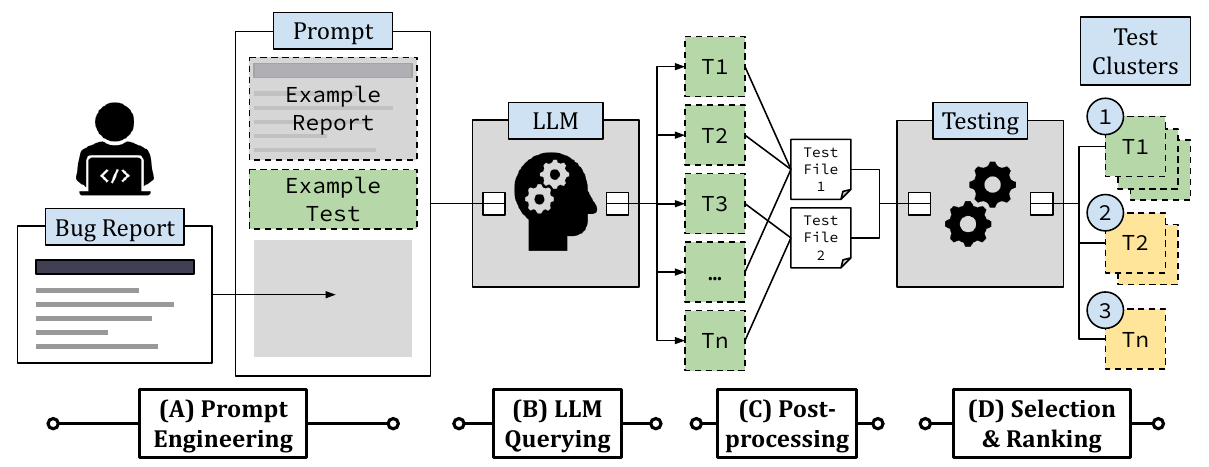}
\caption{Overview of \name\label{fig:overview}}
\end{figure}

\Cref{fig:overview} presents a schematic of our approach, \name. In step (A),
a bug report is used to construct a 'prompt' which conditions an LLM to generate
a bug-reproducing test corresponding to the content of the bug report. This prompt
is used in step (B), where an LLM generates multiple candidate bug-reproducing
tests based on the prompt. However, showing all such results to a developer
would be overwhelming, so tests are executed. First, in step (C), generated tests
are injected into the existing test suite so that they can be executed; then, in
step (D), execution results are used to filter out and rank generated tests so that
developers only need to inspect the most promising generated tests. In the remainder
of this section, we describe each step in greater detail using the example provided in \Cref{tab:ex-report-math-63}.

\subsection{Prompt Engineering}

LLMs are, at the core, large autocomplete neural networks: prior work has found that different ways of `asking' the LLM to solve a problem will lead
to significantly varying levels of performance~\cite{Kojima2022Zreason}. Finding the best query to accomplish the given task is 
known as \textit{prompt engineering}~\cite{reynolds2021prompt}.

To make an LLM generate a test method from a given bug report, we construct 
a Markdown document from the report, which is used as the prompt. For example, 
consider the example in Listing~\ref{lst:ex-prompt-math-63}, which is a 
Markdown document constructed from the bug report shown in 
\Cref{tab:ex-report-math-63}. \name adds a few distinctive parts to the Markdown document: the command ``Provide a self-contained example that 
reproduces this issue'', the start of a block of code in Markdown, (i.e., 
\texttt{\textasciigrave\textasciigrave\textasciigrave}), and finally the partial code snippet \code{public void test} 
which induces the LLM to write a test method.
\begin{table}[h!]
    \centering
    \caption{\label{tab:ex-report-math-63}Example bug report (Defects4J Math-63).}
    \begin{tabular}{@{}lp{0.38\textwidth}@{}}
    \toprule
    Issue No.   & MATH-370\tablefootnote{\url{https://issues.apache.org/jira/browse/MATH-370}}\\ \midrule
    Title     & NaN in ``equals'' methods\\ \midrule
    Description & \begin{tabular}[c]{@{}l@{}}In ``MathUtils'', some ``equals'' methods will return \\
        true if both argument are NaN. Unless I'm mistaken, \\ this contradicts the IEEE standard.\\ 
        If nobody objects, I'm going to make the changes.\end{tabular} \\ \bottomrule
    \end{tabular}
\end{table}

\begin{lstlisting}[basicstyle=\footnotesize\ttfamily,
    columns=flexible,
    breaklines=true,
    caption={Example prompt without examples.},
    label={lst:ex-prompt-math-63},]
# NaN in "equals" methods
## Description
In "MathUtils", some "equals" methods will return true if both argument are NaN.  
Unless I'm mistaken, this contradicts the IEEE standard.
If nobody objects, I'm going to make the changes.

## Reproduction
>Provide a self-contained example that reproduces this issue.
```
public void test
\end{lstlisting}

We evaluate a range of variations of this basic prompt. Brown et al.~\cite{
brown2020language} report that LLMs benefit from question-answer examples 
provided in the prompt. In our case, this means providing examples of bug 
reports (questions) and the corresponding bug reproducing tests (answers). With 
this in mind, we experiment with a varying number of examples, to see whether 
adding more examples, and whether having examples from within the same project 
or from other projects, significantly influences performance.

As there is no restriction to the prompt format, we also experiment with 
providing stack traces for crash bugs (to simulate situations where a stack 
trace was provided), or providing constructors of the faulty class
(to simulate situations where the location of the bug is known).

Our specific template format makes it highly unlikely that prompts we generate
exist verbatim within the LLM training data. Further, most reports in practice
are only connected to the bug-revealing test via a chain of references. As 
such, our format partly mitigates data leakage concerns, among other steps 
taken to limit this threat described later in the manuscript. Finally, when
using chat-optimized models such as ChatGPT in our experiments, we add examples
within the prompt, although in our experiments we needed to update the prompt format 
as the model was updated, as discussed in \Cref{sec:cgpt_change_results}.

\subsection{Querying an LLM}

Using the generated prompt, \name queries the LLM to predict the tokens that 
would follow the prompt. Due to the nature of the prompt, it is likely to 
generate a test method, especially as our prompt ends with the sequence 
\code{public void test}. We ensure that the result only spans the test method 
by accepting tokens until the first occurrence of the string \code{\`\`\`}, which 
indicates the end of the code block in Markdown.

It is known that LLMs yield inferior results when performing completely greedy decoding (i.e., decoding strictly based on the most likely next token)~\cite{brown2020language}: they perform better when they are doing weighted random 
sampling, a behavior modulated by the \textit{temperature} parameter. Following 
prior work, we set our temperature to 0.7~\cite{brown2020language}, which allows
the LLM to generate multiple distinct tests based on the exact same prompt, although
we evaluate the effect of temperature in \Cref{sec:temp_perf_results} and \Cref{sec:temp_ranking_results}.
We take the 
approach of generating multiple candidate reproducing tests, then using their 
characteristics to identify how likely it is that the bug is actually 
reproduced.

An example output from the LLM given the prompt in 
Listing~\ref{lst:ex-prompt-math-63} is shown in 
Listing~\ref{lst:ex-test-success-math-63}: at this point, the outputs from the LLM typically cannot be compiled on their own, and need other constructs such as import statements. We next present how \name integrates a generated test into the existing test suite to make it executable.

\begin{lstlisting}[basicstyle=\footnotesize\ttfamily,
    columns=flexible,
    breaklines=true,
    language=java,
    caption={Example LLM result from the bug report described in Table~\ref{tab:ex-report-math-63}.},
    label={lst:ex-test-success-math-63},]
public void testEquals() {
    assertFalse(MathUtils.equals(Double.NaN, Double.NaN));
    assertFalse(MathUtils.equals(Float.NaN, Float.NaN));
}
\end{lstlisting}

\subsection{Test Postprocessing}

We first describe how \name injects a test method into an existing suite, then how \name resolves the remaining unmet dependencies.

\subsubsection{Injecting a test into a suitable test class}
If a developer finds a test method in a bug report, they will likely insert it 
into a test class which will provide the required context for the test method 
(such as the required dependencies). For example, for the bug in our running 
example, the developers added a reproducing test to the \code{MathUtilsTest} class, where 
most of the required dependencies are already imported, including the focal 
class, \code{MathUtils}. Thus, it is natural to also inject LLM-generated tests 
into existing test classes, as this matches developer workflow, while resolving 
a significant number of initially unmet dependencies.

\begin{lstlisting}[basicstyle=\footnotesize\ttfamily,
    columns=flexible,
    breaklines=true,
    language=java,
    caption={Target test class to which the test in Listing~\ref{lst:ex-test-success-math-63} is injected.},
    label={lst:ex-injected-class-math-63},]
public final class MathUtilsTest extends TestCase {
    ...
    public void testArrayEquals() {
        assertFalse(MathUtils.equals(new double[] { 1d }, null));
        assertTrue(MathUtils.equals(new double[] {
            Double.NaN, Double.POSITIVE_INFINITY,
        ...
\end{lstlisting}

To find the best test class to inject our test methods into, we find the test class
that is \emph{lexically} most similar to the generated test 
(Algorithm~\ref{alg:1-test-postprocessing}, line 1). The intuition is that, 
if a test method belongs to a test class, the test method likely uses similar 
methods and classes, and is thus lexically related, to other tests from that 
test class. Formally, we assign a matching score for each test class based on \Cref{eq:test2class}:
\begin{equation}
    sim_{c_i} = |T_{t} \cap T_{c_i}| / |T_{t}| 
    \label{eq:test2class}
\end{equation}
where $T_t$ and $T_{c_i}$ are the set of tokens in the generated test method 
and the $i$th test class, respectively. 
As an example, Listing~\ref{lst:ex-injected-class-math-63} shows the 
key statements of the \code{MathUtilsTest} class. Here, the test class contains 
similar method invocations and constants with those used by the LLM-generated 
test in Listing~\ref{lst:ex-test-success-math-63}, particularly in
lines 4 and 6.

As a sanity check, we inject ground-truth developer-added bug reproducing tests 
from the Math and Lang projects of the Defects4J benchmark, and check if they 
execute normally based on Algorithm~\ref{alg:1-test-postprocessing}. We find 
execution proceeds as usual for 89\% of the time, suggesting that the algorithm 
reasonably finds environments in which tests can be executed.

\RestyleAlgo{ruled} 
\SetKwComment{Comment}{/* }{ */}
\begin{algorithm}[t]
\footnotesize
\caption{Test Postprocessing}\label{alg:1-test-postprocessing}
\KwIn{A test method $tm$; Test suite $\mathcal{T}$ of SUT; source code files $\mathcal{S}$ of SUT;}
\KwOut{Updated test suite $\mathcal{T}'$}

$c_{best} \leftarrow$ \codealgo{findBestMatchingClass\(}$tm$, $\mathcal{T}$\codealgo{\)}\;
$deps \leftarrow$ \codealgo{getDependencies\(}$tm$\codealgo{\)}\;
$needed\_deps \leftarrow$ \codealgo{getUnresolved\(}$deps$, $c_{best}$\codealgo{\)}\;
$new\_imports \leftarrow$ \codealgo{set()}\;    
\For{$dep$ in $needed\_deps$}{
    $target \leftarrow$ \codealgo{findClassDef\(}$dep$, $\mathcal{S}$\codealgo{\)}\;
    \eIf{$target$ is \codealgo{null}}{
        $new\_imports$\codealgo{.add\(findMostCommonImport\(}$dep, \mathcal{S}, \mathcal{T}$\codealgo{\)\)}\;
    }{
        $new\_imports$\codealgo{.add\(}$target$\codealgo{\)}\;
    }
}
$\mathcal{T}' \leftarrow$ \codealgo{injectTest\(}$tm$, $c_{best}$, $\mathcal{T}$\codealgo{\)}\;
$\mathcal{T}' \leftarrow$ \codealgo{injectDependencies\(}$new\_imports, c_{best}, \mathcal{T}'$\codealgo{\)}\;
\end{algorithm}

\subsubsection{Resolving remaining dependencies}
Although many dependency issues are resolved by placing the test in the right 
class, the test may introduce new constructs that need to be imported.
To handle these cases, \name heuristically infers packages to import. 

Line $2$ to $10$ in Algorithm~\ref{alg:1-test-postprocessing} describe the 
dependency resolving process of \name. First, \name parses the generated test 
method and identifies variable types and referenced class 
names/constructors/exceptions. \name then filters ``already imported'' class 
names by lexically matching names to existing import statements in the test 
class (Line $3$). 

As a result of this process, we may find types that are not resolved within the
test class. \name first attempts to find public classes with the identified  
name of the type; if there is exactly one such file, the classpath to the 
identified class is derived (Line $7$), and an import statement is added 
(Line $11$). However, either no or multiple matching classes may exist. In both 
cases, \name looks for \code{import} statements ending with the target class 
name within the project (e.g., when searching for \code{MathUtils}, \name looks 
for \code{import .*MathUtils;}). \name selects the most common import statement 
across all project source code files. Additionally, a few rules 
ensure assertion statements are properly imported, even when there are no 
appropriate imports within the project itself.

Our postprocessing pipeline does not guarantee compilation in all cases, but 
the heuristics used by \name are capable of resolving most of the unhandled 
dependencies of a raw test method. After going through the postprocessing 
steps, \name executes the tests to identify candidate bug reproducing 
tests. 

\subsection{Selection and Ranking}
\label{sec:sel_n_rank}

A test is a Bug Reproducing Test (BRT) if and only if the test fails due to the 
bug specified in the report. A \emph{necessary} condition for a test generated 
by \name to be a BRT is that the test compiles and fails in the buggy program: we call such tests FIB (Fail In the Buggy program) tests. 
However, not all FIB tests are BRTs, making it difficult to tell whether bug 
reproduction has succeeded or not. This is one factor that separates us from 
crash reproduction work~\cite{Soltani2020aa}, as crash reproduction techniques can confirm whether the bug has been reproduced by comparing the stack traces at the time of crash. 
On the other hand, it is imprudent to present all generated FIB tests 
to developers, as asking developers to iterate over multiple solutions is 
generally undesirable~\cite{Kochhar2016FLSurvey, Noller2022APRSurvey}.
As such, \name attempts to decide when to suggest a test and, if so, which test 
to suggest, using several patterns we observe to be correlated to successful 
bug reproduction. 

\SetKwComment{Comment}{/* }{ */}
\SetKwProg{Fn}{Function}{ is}{end}
\begin{algorithm}[t]
    \footnotesize
\caption{Test Selection and Ranking}\label{alg:2-test-ranking}
\KwIn{Pairs of modified test suites and injected test methods $\mathcal{S_{T'}}$; target program with bug $P_{b}$;  bug report $BR$; agreement threshold $Thr$;}
\KwOut{Ordered list of ranked tests $ranking$; }
$FIB \leftarrow $ \codealgo{set()}\; 
\For{$(\mathcal{T}', tm_i) \in \mathcal{S_{T'}}$}{
    $r \leftarrow$ \codealgo{executeTest(}$\mathcal{T'}, P_{b}$\codealgo{)}\;
    \If{\codealgo{hasNoCompileError(}$r$\codealgo{) && isFailed(}$tm_i, r$\codealgo{)}}{
        $FIB$\codealgo{.add(}$(tm_i, r)$\codealgo{)}\;
    }
}

$clusters\leftarrow$\codealgo{clusterByFailureOutputs(}$FIB$\codealgo{)}\;
$output\_clus\_size \leftarrow clusters$\codealgo{.map(size)}\;
$max\_output\_clus\_size \leftarrow $\codealgo{max(}$output\_clus\_size$\codealgo{)}\;
\If{$max\_output\_clus\_size \leq Thr$}{
    \Return{\codealgo{list()}}\;
}
$FIB_{uniq} \leftarrow$ \codealgo{removeSyntacticEquivalents(}$FIB$\codealgo{)}\;

$br\_output\_match \leftarrow clusters$\codealgo{.map(matchOutputWithReport(}$BR$\codealgo{))}\;
$br\_test\_match \leftarrow FIB_{uniq}$\codealgo{.map(matchTestWithReport(}$BR$\codealgo{))}\;
$tok\_cnts \leftarrow FIB_{uniq}$\codealgo{.map(countTokens)}\;
$ranking \leftarrow $\codealgo{list()}\; 
$clusters \leftarrow clusters$\codealgo{.sortBy(}$br\_output\_match, output\_clus\_size,tok\_cnts$\codealgo{)}\;
\For{$clus \in clusters$}{
    $clus \leftarrow clus$\codealgo{.sortBy(}$br\_test\_match, tok\_cnts$\codealgo{)}\;
}
\For{$i=0; i<$\codealgo{max(}$output\_clus\_size$\codealgo{)}$;i\gets i+1$}{
\For{$clus \in clusters$}{
    \If{$i<clus$\codealgo{.length()}}{$ranking$\codealgo{.push(}$clus$\codealgo{[i])}}
}
}
\Return{$ranking$}\;

\end{algorithm}

Algorithm~\ref{alg:2-test-ranking} outlines how \name decides whether to 
present results and, if so, how to rank the generated tests. In Line 1-10, \name 
first decides whether to show the developer any results at all (selection). We 
group the FIB tests that exhibit the same failure output (the same error type 
and error message) and look at the number of the tests in the same group (the 
$max\_output\_clus\_size$ in Line 8). This is based on the intuition that, if 
multiple tests show similar failure behavior, then it is likely that the LLM is 
`confident' in its predictions as its independent predictions `agree' with each other, and 
there is a good chance that bug reproduction has succeeded. This is a well-known
and universal property of LLMs, also explored by Wang et al.~\cite{wang2023selfconsistency}
and verified in our experiments. \name can be 
configured to only show results when there is significant agreement in the 
output (setting the agreement threshold $Thr$ high) or show more exploratory 
results (setting $Thr$ low).

Once it decides to show its results, \name relies on three heuristics to rank 
generated tests, in ascending order of discriminative strength.
First, tests are likely to be bug reproducing if the fail message and/or the 
test code shows the behavior (exceptions or output values) observed and 
mentioned in the bug report. While this heuristic is precise, its decisions are not very discriminative, as tests can only be divided into groups of 
`contained' versus `not contained'. Next, we look at the `agreement' between 
generated tests by looking at output cluster size ($output\_clus\_size$), which 
represents the `consensus' of the LLM generations. Finally, \name prioritizes 
based on test length (as shorter tests are easier to understand), which is the 
finest-grained signal. We first leave only syntactically unique tests 
(Line 11), then sort output clusters and tests within those clusters 
using the heuristics above (Lines 16 and 18). 

As tests with the same failure output are similar to each other, we expect 
that, if one test from a cluster is not BRT, the rest from the same cluster are 
likely not BRT as well. Hence, \name shows tests from a diverse array of 
clusters. For each $i$th iteration in Line 19-22, the $i$th ranked 
test from each cluster is selected and added to the list.

\begin{table*}[]
    \centering
    \begin{tabular}{@{}cccccccc@{}} 
    \toprule 
    Organization                & LLM              & Accessible          & Size & 
    Downloadable & Code Model & Release Year & Chat \\ 
    \midrule 
    \multirow{4}{*}{OpenAI}     & code-davinci-002 (Codex)       & Not Since Mar. 2023 & 176B?   & 
    No           & Yes        & 2021         & No             \\ 
                                & text-davinci-003 & Yes                 & 176B?   & 
    No           & No         & 2022         & No             \\ 
                                & gpt-3.5-turbo-0301    & Yes                 & ?        & 
    No           & No         & 2023         & Yes            \\ 
                                & gpt-3.5-turbo-0613    & Yes                 & ?        & 
    No           & No         & 2023         & Yes            \\ 
    \multirow{2}{*}{BigScience} & Bloom            & Yes                 & 176B    & 
    Yes          & No         & 2022         & No             \\ 
                                & BloomZ           & Yes                 & 176B    & 
    Yes          & No         & 2022         & No             \\ 
    Meta       & Incoder          & Yes                 & (1, 6)B     & 
    Yes          & Yes        & 2022         & No             \\ 
    Salesforce                  & CodeGen2      & Yes                 & (1, 3.7, 7, 16)B     & 
    Yes          & Yes        & 2023         & No             \\ 
    \multirow{3}{*}{BigCode}    & StarCoder        & Yes                 & 15B    & 
    Yes          & Yes        & 2023         & No             \\ 
                                & StarCoderBase    & Yes                 & 15B    & 
    Yes          & Yes        & 2023         & No             \\ 
                                & StarCoderPlus    & Yes                 & 15B    & 
    Yes          & Yes        & 2023         & Yes             \\ 
    \bottomrule 
    \end{tabular} 
    \caption{LLMs used in our evaluation.}
    \label{tab:llm_info}
\end{table*}

\section{Evaluation}
\label{sec:evaluation}
This section provides evaluation details for our experiments.

\subsection{Dataset}
\label{sec:dataset}

To evaluate \name, and its performance when using different LLMs, over a large
number of real-world bugs, we employ the Defects4J v2.0 dataset, which is a collection
of Java bugs with bug reports\footnote{Except for the Chart project, 
for which only 8 bugs have reports} collected from open-source repositories. Consequently,
the dataset provides a convenient means of evaluating how many bugs \name could reproduce
over different projects. While there are 814 bugs in the Defects4J v2.0 dataset, 58 bugs
were excluded as upon inspection they were not well mapped to the bug; furthermore, six bugs
were excluded as they had a different directory structure between the buggy and fixed versions,
making it difficult to consistently apply our evaluation pipeline. Thus, we evaluated over
\textbf{750} bugs that had a good report-bug matching, and for which there was not substantial
refactoring between the buggy and fixed versions. To compare against the state-of-the-art crash reproduction
technique EvoCrash~\cite{Soltani2018aa}, 
we used the 60 bugs in Defects4J that were included in the JCrashPack~\cite{soltani2020benchmark}
dataset.

The Defects4J dataset is likely in most code-based LLM training data, as was
argued by the recent work of Lee et al.~\cite{lee2023github},
even if the prompt format we use is 
unlikely to have appeared verbatim in the data. Indeed, Lee et al. note
that specifically for StarCoder, many of the Defects4J bug-reproducing tests are
included in the StarCoder pretraining data. This threats conclusions derived from
Defects4J, as it is unclear whether any results are due to LLM memorization or are
genuinely likely to be reproducible by practitioners in their own projects.
To mitigate such concerns, from 
17 GitHub repositories\footnote{These repositories have been manually chosen 
from either Defects4J projects that are on GitHub and open to new issues, or Java projects that have been modified since 10th July 2022 with at least 100 or more stars, as of 1st of August 2022. A list of the 17 repositories is available in the artifact of our prior work~\cite{Kang2023LIBRO}.} that use JUnit, we gather 581 Pull Requests (PR) created after the Codex training data cutoff point, ensuring that 
this dataset could not have been used to train Codex. We further 
check if a PR adds any test to the project (435 left after discarding  
non-test-introducing ones), and filter out the PRs that are not merged to the 
main branch or associated with multiple issues - this leaves 84 PRs.

As a final step, we verified that each PR added a bug-reproducing test by
verifying that the developer-added test in the PR failed on the pre-merge commit
of the project, but passed on the post-merge commit. PRs that failed to introduce
any such tests were discarded. This left us with a final dataset of \textbf{31}
bugs which could be reproduced, and their corresponding bug reports. In the
remainder of this paper, we refer to this dataset as the GHRB (GitHub Recent Bugs)
dataset; we use this dataset to verify that the results we obtain for Defects4J
are not simply due to LLM verbatim memorization of training data.

\subsection{Metrics} 
The goal of \name is to generate Bug Reproducing Tests (BRTs) for each bug.
We treat a generated test as a BRT if and only if it (i) fails on the initially
buggy version of the code and (ii) passes on the fixed version of the code.
For a bug to be \textit{reproduced} by \name, one or more BRTs must be
generated for the bug. For each setting and technique, the number of BRTs is counted.

To maximize the realism of our evaluation, when using the Defects4J benchmark,
we use the \texttt{PRE\_FIX\_REVISION} and \texttt{POST\_FIX\_REVISION} versions
of the code, which reflect the unaltered state of the project when a bug was
reported/fixed. When using the GHRB suite of bugs, we use the commits before and
after a pull request as buggy and fixed versions.

As a baseline, we compare against the EvoCrash~\cite{Soltani2018aa} technique.
By default, EvoCrash will only check whether the crash stack is reproduced, instead
of verifying over the fixed version. To make comparison consistent, we apply our
definition of BRT, and execute EvoCrash tests on the buggy and fixed versions
to determine success.

As \name not only produces BRTs, but also selects results and BRTs, evaluation
metrics are required for this aspect as well. For result selection, we use the
ROC-AUC metric, which evaluates the performance of a classifier regardless of
the specific threshold used. To evaluate the rankings, we use the $acc@n$ and
$wef$ metrics. $acc@n$ measures the number of bugs for which \name could successfully
rank a BRT within the top $n$ suggestions. $wef$ measures the amount of wasted
developer effort when inspecting the suggested tests from \name; a variant of it,
$wef@n$, measures the wasted effort when inspecting the top $n$ candidates.

\subsection{Environment} 

All test execution experiments were performed on a machine running Ubuntu 18.04.6 LTS, 
with 32GB of RAM and Intel(R) Core(TM) i7-7700 CPU @ 3.60GHz CPU. 
Meanwhile, we set up a server for hosting open-source LLM inference, which ran Ubuntu 20.04.6 LTS and was equipped with 16 Intel(R) Xeon(R) Gold 5222 CPU @ 3.80GHz CPUs and four NVIDIA RTX 3090 GPUs, with a total VRAM of 96GB. This was enough to run the 15B StarCoder LLM which has 8,000 context length without weight quantization. The server was implemented to behave similarly to the OpenAI API, with configurable LLMs being loaded as needed to run the experiments. 

Regarding the LLMs we use, we present their characteristics in Table~\ref{tab:llm_info}. For the closed-source LLMs from OpenAI, we access them via the public API, and evaluate the performance of the \code{gpt-3.5-turbo-0301} and \code{gpt-3.5-turbo-0613} chat-based models as well as the autocompletion-based \code{text-davinci-003} model, and compare the bug reproduction results with the results of using the \code{code-davinci-002} (Codex) model from our prior work~\cite{Kang2023LIBRO}. For the open-source LLMs, we evaluate the BLOOM~\cite{workshop2023bloom}/BLOOMZ~\cite{muennighoff2023crosslingual} family of LLMs, which were the largest open-source LLMs at the time of experimentation; thus, we decided to experiment with them, even though they were not specifically trained on code. As our infrastructure could not host the BLOOM LLMs, we used the HuggingFace inference API to conduct our experiments. All the other LLMs that we used were trained on code. The InCoder~\cite{fried2023incoder} family of LLMs consists of two LLMs of different sizes, and are code-based LLMs trained by what is now Meta, with the largest model having 6 billion parameters. The CodeGen2~\cite{nijkamp2023codegen2} family of LLMs consisting of four LLMs of different sizes, was trained by Salesforce, with the largest model having 16 billion parameters. Finally, there were three variants of the StarCoder~\cite{li2023starcoder} model; the initial StarCoderBase model which was trained on a large corpus (more than one trillion tokens) of source code, the StarCoder model which is a fine-tuned version of StarCoderBase on Python code, and the StarCoderPlus model which was further fine-tuned from the StarCoder model so that it could operate in a chat format. We utilize the variants of the open-source models to discuss the effect of training data and model size on bug reproduction performance.

For all models, we set the sampling temperature to 0.7 (except in \Cref{sec:temp_perf_results} and \Cref{sec:temp_ranking_results} where we explicitly experiment with the temperature parameter), and the maximum 
number of generated tokens to 256. When comparing settings and LLMs in Defects4J, by default we sample 10 tests (denoted by n=10), due to computational constraints; however, for the
Codex (code-davinci-002) and StarCoder LLM, we experiment with sampling 50 tests as well. For the GHRB dataset, due to the small number of bugs, we consistently evaluate with n=50 tests.
We script our experiments using Python 3.9, and parse 
Java files with the \texttt{javalang} library~\cite{c2nes2022javalang}.
Our tool is public\footnote{\url{https://github.com/coinse/libro}}, and the replication package for our journal extension is also available as well\footnote{\url{https://github.com/coinse/libro-journal-artifact}}.

\section{Research Questions}
\label{sec:rqs}
This section presents our research questions. In RQ1-3, we present the effectiveness
of \name when using the Codex (\texttt{code-davinci-002}) LLM from OpenAI. In RQ3 and RQ4,
we use different LLMs and LLM settings for \name and compare their efficacy.

\subsection{RQ1: Efficacy}

In RQ1, the bug-reproducing performance of \name when using Codex is evaluated and compared
to baselines, based on the Defects4J dataset.

\begin{itemize}
    \item \textbf{RQ1-1: How many bug reproducing tests can \name generate?} We present how many bugs \name could reproduce based on different prompts, when using the best-performing LLM, \texttt{code-davinci-002}.
    \item \textbf{RQ1-2: How does \name compare to other techniques?} We compare against baselines to further demonstrate the performance of \name. As there are no general bug reproduction techniques that we are aware of, we compare against the state-of-the-art crash reproduction technique EvoCrash. Furthermore, we implement and compare with a `Copy\&Paste' baseline which directly uses code snippets from the bug report, extracted via infoZilla~\cite{Premraj2008infoZilla} or by the HTML \code{<pre>} tag, as bug-reproducing tests.
\end{itemize}

\subsection{RQ2: Efficiency} 

RQ2 inspects various aspects of the efficiency of \name, so that practitioners
may estimate the costs of deploying \name on new projects.

\begin{itemize}
    \item \textbf{RQ2-1: How many LLM queries are required?} The number of queries to Codex required to reach a certain number of bugs reproduced on the Defects4J dataset is evaluated.
    \item \textbf{RQ2-2: How much time does \name need?} The amount of computation time used at each stage of the \name pipeline is evaluated.
    \item \textbf{RQ2-3: How many tests should the developer inspect?} To estimate the degree of developer effort needed to use \name, the number of bugs for which tests would be suggested, and the number of tests developers would need to inspect, is evaluated.
\end{itemize}

\subsection{RQ3: Practicality}

RQ3 investigates how well \name with Codex generalizes, by evaluating it on the
GHRB dataset instead of the Defects4J dataset, which may be part of Codex training data.

\begin{itemize}
    \item \textbf{RQ3-1: How often can \name reproduce bugs in the wild?} The number of bugs that can be reproduced by \name on the GHRB dataset is evaluated, mitigating data leakage risks.
    \item \textbf{RQ3-2: How reliable are the selection and ranking techniques of \name?} Whether the selection and ranking algorithm work on an unseen dataset is evaluated, demonstrating that the pipeline captures generalizable properties of Codex.
\end{itemize}

\subsection{RQ4: Sensitivity to LLMs}

With RQ4, we present the results of a large-scale study involving 15 LLMs, and present how the performance of \name changes as the underlying LLM varies.

\begin{itemize}
    \item \textbf{RQ4-1: What is the best LLM to use given GPU limits?} We present the results of evaluating \name with multiple different LLMs, and propose which LLM performs best when GPU memory is restricted.
    \item \textbf{RQ4-2: How well do the LLMs perform on holdout bugs?} We evaluate \name with different LLMs on the holdout GHRB bug dataset to see whether the LLMs used by \name are simply relying on dataset memorization.
    \item \textbf{RQ4-3: How has ChatGPT's behavior changed over time?} We compare the performance of \name when different ChatGPT models are used as the LLM, similarly to Chen et al.~\cite{chen2023chatgptchange}, and analyze why such a change occurred.
    \item \textbf{RQ4-4: How does LLM size influence performance within the same LLM family?} By comparing LLMs from the same family, which were trained with the same data and techniques with the only difference being the model size, we compare how \name performance changes as LLM size is increased.
    \item \textbf{RQ4-5: How does LLM sampling temperature influence performance?} LLMs have a `temperature' parameter that controls the degree of randomness in the outputs of LLMs. The influence the temperature parameter has on the performance of \name-StarCoder is evaluated.
\end{itemize}

\subsection{RQ5: Sensitivity to LLM Self-Consistency}

The selection and ranking algorithm of \name, while developed independently, shares the same intuition as the phenomenon in LLMs known as self-consistency~\cite{wang2023selfconsistency}. Consequently, the selection and ranking algorithm of \name is a good way of evaluating how consistent the self-consistency phenomenon is over different LLMs.

\begin{itemize}
    \item \textbf{RQ5-1: Does the choice of LLM influence the performance of \name postprocessing?} Different LLMs may have different self-consistency characteristics, and consequently the selection and ranking algorithm of \name may show different performance. To evaluate this, we use the results of multiple LLMs from RQ4 and evaluate how well our selection and ranking algorithm works.
    \item \textbf{RQ5-2: Does the sampling temperature of LLM change the performance of \name postprocessing?} The sampling temperature of LLMs has an even more direct relationship to self-consistency -- when the temperature is low, more results will be similar and vice versa. To this end, we evaluate what influence sampling temperature has on our selection and ranking algorithm.
\end{itemize}

\section{Experimental Results}
\label{sec:results}

\subsection{RQ1. How effective is \name?}

\subsubsection{RQ1-1}
\Cref{tab:d4j_per_prompt} shows which prompt/information settings work best, where
$n=N$ means we queried the \texttt{code-davinci-002} $N$ times for reproducing tests.
When using examples from the source project, we use the oldest tests available 
within that project; otherwise, we use two handpicked report-test pairs 
(Time-24, Lang-1) throughout all projects. We find that providing constructors 
(\textit{\`{a} la} AthenaTest~\cite{tufano2020unit}) does not help 
significantly, but adding stack traces does help reproduce crash bugs, 
indicating that \name can benefit from using the stack information to replicate 
issues more accurately. Interestingly, adding within-project examples shows 
poorer performance: inspection of these cases has revealed that, in such cases, 
\name simply copied the provided example even when it should not have, leading 
to lower performance. We also find that the number of examples makes a 
significant difference (two-example n=10 values are sampled from n=50 results 
from the default setting), confirming the existing finding that adding examples helps improve
performance. In 
turn, the number of examples seems to matter less than the number of times the 
LLM is queried, as we further explore in RQ2-1. As the two-example n=50 
setting shows the best performance, we use it as the default setting 
throughout the rest of the paper.

\begin{table}[ht]
    \centering
    \caption{Reproduction performance for different prompts\label{tab:d4j_per_prompt}}
    \scalebox{0.9}{
    \begin{tabular}{lcc}
    \toprule
    Setting & reproduced & FIB \\\midrule
    No Example (n=10) & 124 & 440 \\
    One Example (n=10) & 166 & 417 \\
    One Example from Source Project (n=10) & 152 & 455 \\
    One Example with Constructor Info (n=10) & 167 & 430 \\
    Two Examples (n=10, 5th percentile) & 161 & 386 \\
    Two Examples (n=10, median) & 173 & 409 \\
    Two Examples (n=10, 95th percentile) & 184 & 429 \\
    Two Examples (n=50) & \textbf{251} & \textbf{570} \\\midrule
    One Example, Crash Bugs (n=10) & 69 & 153 \\
    One Example with Stack, Crash Bugs (n=10) & 84 & 155 \\
    \bottomrule
    \end{tabular}
    }
\end{table} 

Under the two-example n=50 setting, we find that overall \textbf{251} 
bugs, or 33.5\% of 750 studied Defects4J bugs, are reproduced by \name. 
\Cref{tab:d4j_per_proj} presents a breakdown of the performance per project. 
While there is at least one bug reproduced for every project, the proportion
of bugs reproduced can vary significantly. For example, \name reproduces a 
small number of bugs in the Closure project, which is known to have a unique 
test structure~\cite{Martinez2016AutomaticRO}. On the other hand, the 
performance is stronger for the Lang or Jsoup projects, whose tests are 
generally self-contained and simple. Additionally, we find that the average
length of the generated test body is about 6.5 lines (excluding comments
and whitespace characters), indicating \name is capable of writing meaningfully long
tests.

\begin{table}[ht]
    \centering
    \caption{Bug reproduction per project in Defects4J:\protect\\
    x/y means x reproduced out of y bugs\label{tab:d4j_per_proj}}
    \scalebox{1.0}{
    \begin{tabular}{lr|lr}
    \toprule
    Project & rep/total & Project & rep/total\\\midrule
    Chart & 5/7 & JacksonDatabind & 30/107 \\
    Cli & 14/29 & JacksonXml & 2/6 \\
    Closure & 2/172 & Jsoup & 56/92\\
    Codec & 10/18 & JxPath & 3/19 \\
    Collections & 1/4 & Lang & 46/63\\
    Compress & 4/46 & Math & 43/104 \\
    Csv & 6/16 & Mockito & 1/13 \\
    Gson & 7/11 & Time & 13/19 \\
    JacksonCore & 8/24 & \textbf{Total} & \textbf{251/750} \\
    \bottomrule
    \end{tabular}
    }
\end{table} 

\begin{tcolorbox}[boxrule=0pt,frame hidden,sharp corners,enhanced,borderline north={1pt}{0pt}{black},borderline south={1pt}{0pt}{black},boxsep=2pt,left=2pt,right=2pt,top=2.5pt,bottom=2pt] 
    \textbf{Answer to RQ1-1:} A large (251) number of bugs can be replicated
    automatically using Codex, with bugs replicated over a diverse group of projects.
    Further, the number of examples in the prompt and the number of generation 
    attempts have a strong effect on performance.
\end{tcolorbox}  

\subsubsection{RQ1-2} 
\begin{figure}[h!]
    \centering
    \begin{subfigure}{0.15\textwidth}
        \centering
        \includegraphics[width=\linewidth]{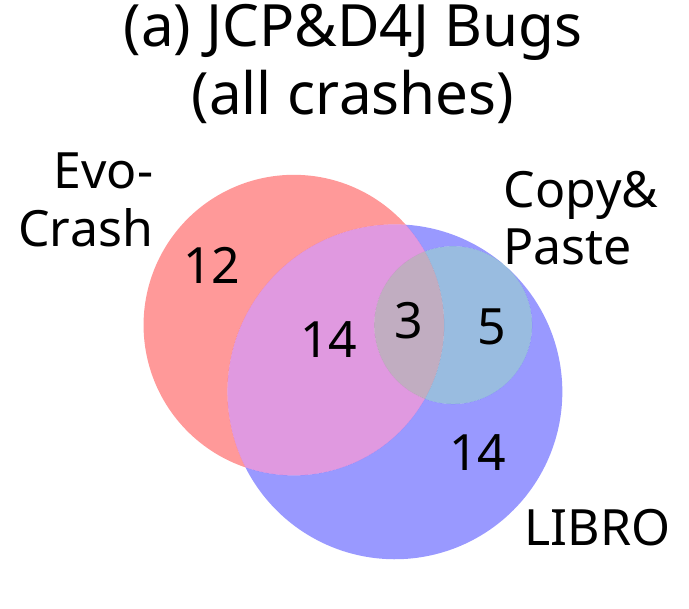}
    \end{subfigure}
    \begin{subfigure}{0.135\textwidth}
        \centering
        \includegraphics[width=\linewidth]{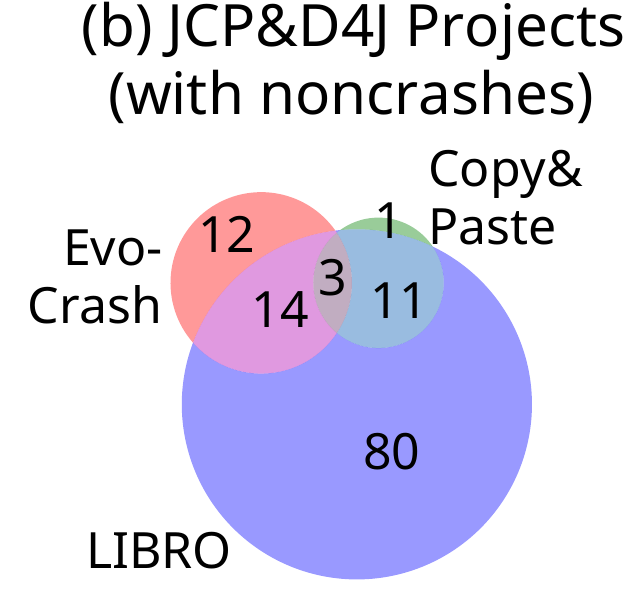}
    \end{subfigure}
    \begin{subfigure}{0.15\textwidth}
        \centering
        \includegraphics[width=\linewidth]{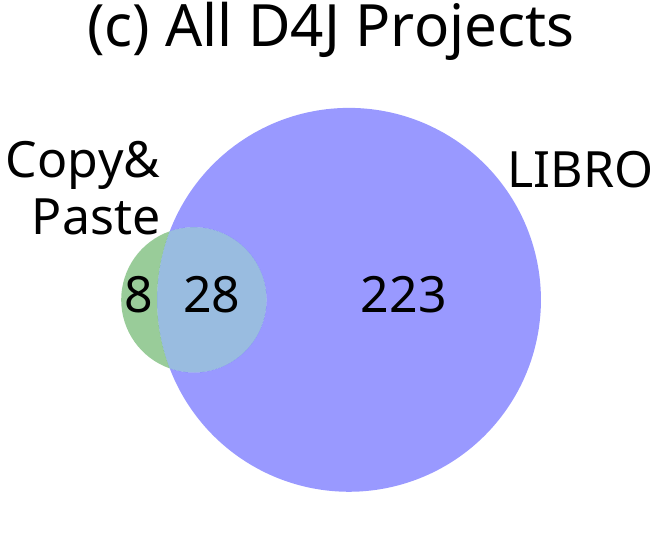}
    \end{subfigure}
    \caption{Baseline comparison on bug reproduction capability}
    \label{fig:baseline-venn}
  \end{figure}

We further compare \name against the 
state-of-the-art crash reproduction technique, EvoCrash, and the `Copy\&Paste 
baseline' that uses code snippets from the bug reports. We present the comparison results in Figure~\ref{fig:baseline-venn}. We find \name 
replicates a large and distinct group of bugs compared to other baselines. 
\name reproduced 91 more unique bugs (19 being crash bugs) than EvoCrash, which 
demonstrates that \name can reproduce non-crash bugs prior work could not 
handle (Fig.~\ref{fig:baseline-venn}(b)). On the other hand, the Copy\&Paste 
baseline shows that, while the BRT is sometimes included in the bug report, the 
report-to-test task is not at all trivial. Interestingly, eight bugs reproduced
by the Copy\&Paste baseline were not reproduced by \name; we find that this is due 
to long tests that exceed the generation length of \name, or due to dependency on 
complex helper functions.

\begin{tcolorbox}[boxrule=0pt,frame hidden,sharp corners,enhanced,borderline north={1pt}{0pt}{black},borderline south={1pt}{0pt}{black},boxsep=2pt,left=2pt,right=2pt,top=2.5pt,bottom=2pt]
    \textbf{Answer to RQ1-2:} \name is capable of replicating a large and distinct group of bugs
    relative to prior work.
\end{tcolorbox}

\subsection{RQ2. How efficient is \name?}

\begin{figure}[h!]
    \centering
    \begin{subfigure}{0.235\textwidth}
        \centering
        \includegraphics[width=\linewidth]{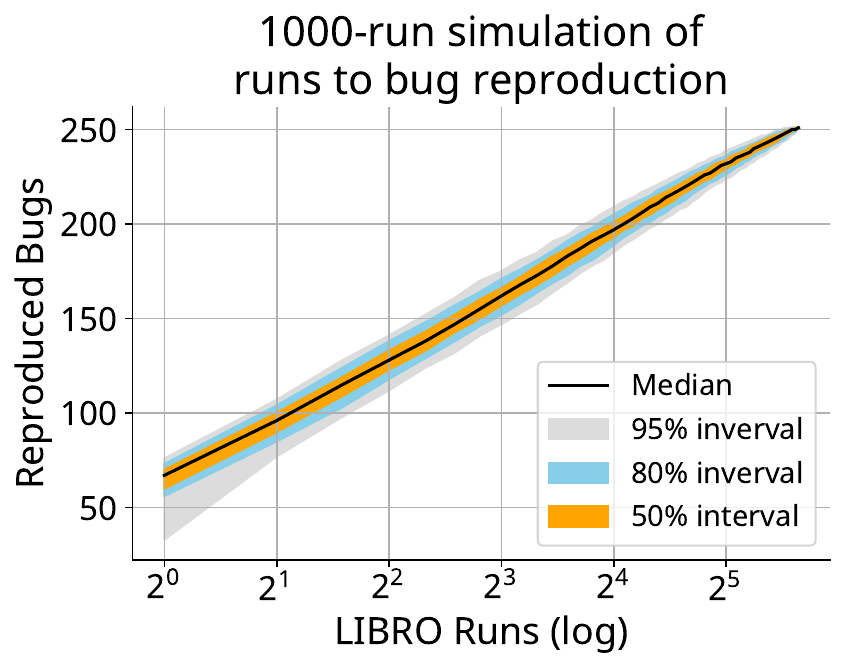}
    \end{subfigure}%
    \hfill
    \begin{subfigure}{0.235\textwidth}
        \centering
        \includegraphics[width=\linewidth]{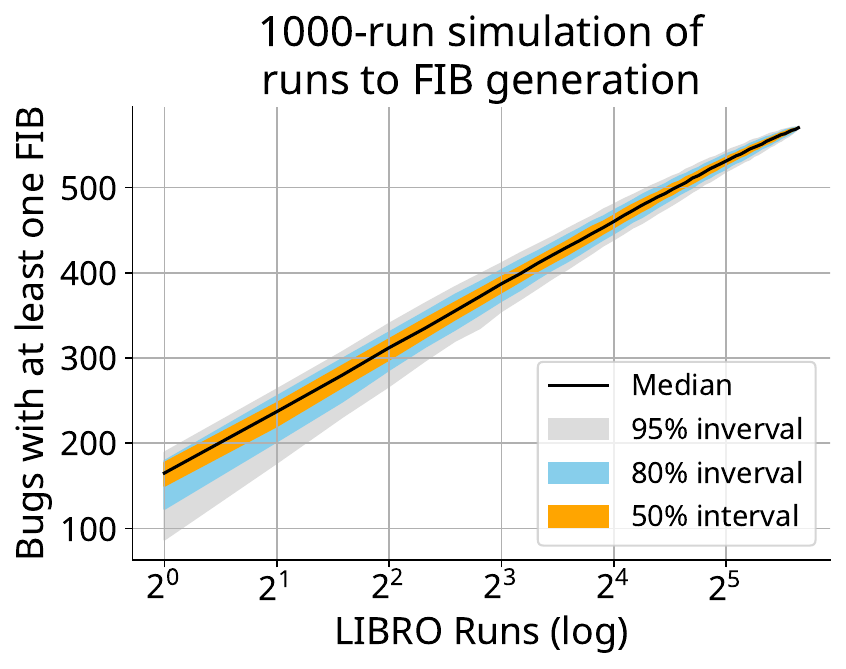}
    \end{subfigure}
    \caption{Generation attempts to performance. Left depicts bugs reproduced as attempts increase, right for FIB}
    \label{fig:run2perf}
  \end{figure}

\subsubsection{RQ2-1} Here, we investigate how many tests must be 
generated to attain a certain bug reproduction performance. To do so, for each 
Defects4J bug, we randomly sample $x$ tests from the 50 generated under the 
default setting, leaving a reduced number of tests per bug. We then check the 
number of bugs reproduced $y$ when using only those sampled tests. We repeat
this process 1,000 times to approximate the distribution.

The results are presented in \Cref{fig:run2perf}. Note that the $x$-axis is in
log scale. Interestingly, we find a logarithmic relation holds between the 
number of test generation attempts and the median bug reproduction performance. 
This suggests that it becomes increasingly difficult, yet stays possible, to 
replicate more bugs by simply generating more tests. As the graph shows no signs
of plateauing, experimenting with an even greater sample of tests may result in 
better bug reproduction results. 

\begin{tcolorbox}[boxrule=0pt,frame hidden,sharp corners,enhanced,borderline north={1pt}{0pt}{black},borderline south={1pt}{0pt}{black},boxsep=2pt,left=2pt,right=2pt,top=2.5pt,bottom=2pt]
    \textbf{Answer to RQ2-1:} The number of bugs reproduced increases
    logarithmically to the number of tests generated, with no sign of performance plateauing.
\end{tcolorbox}  

\begin{table}[ht]
    \centering
    \caption{The time required for the pipeline of \name\label{tab:time_performance}}
    \scalebox{0.9}{
    \begin{tabular}{ccccccc}
    \toprule
     & Prompt & API & Processing & Running & Ranking & Total\\ \midrule
    Single Run & <1 $\mu$s & 5.85s & 1.23s & 4.00s & - & 11.1s\\
    50-test Run & <1 $\mu$s & 292s & 34.8s & 117s & 0.02s & 444s\\
    \bottomrule
    \end{tabular}
    }
\end{table} 

\subsubsection{RQ2-2} We report the time it takes to perform each step of
our pipeline in \Cref{tab:time_performance}. We find API querying takes the 
greatest amount of time, requiring about 5.85 seconds. 
Postprocessing and test executions take 1.23 and 4 seconds per test (when the 
test executes), respectively. Overall, \name took an average of 444 seconds to generate 50 
tests and process them, which is well within the 10-minute search budget often 
used by search-based techniques~\cite{Soltani2020aa}.

\begin{tcolorbox}[boxrule=0pt,frame hidden,sharp corners,enhanced,borderline north={1pt}{0pt}{black},borderline south={1pt}{0pt}{black},boxsep=2pt,left=2pt,right=2pt,top=2.5pt,bottom=2pt]
    \textbf{Answer to RQ2-2:} Our time measurement suggests that \name does not take a significantly
    longer time than other methods to use.
\end{tcolorbox}

\subsubsection{RQ2-3}
\begin{figure}[ht]
    \centering
    \begin{subfigure}{0.232\textwidth}
        \centering
        \includegraphics[width=\linewidth]{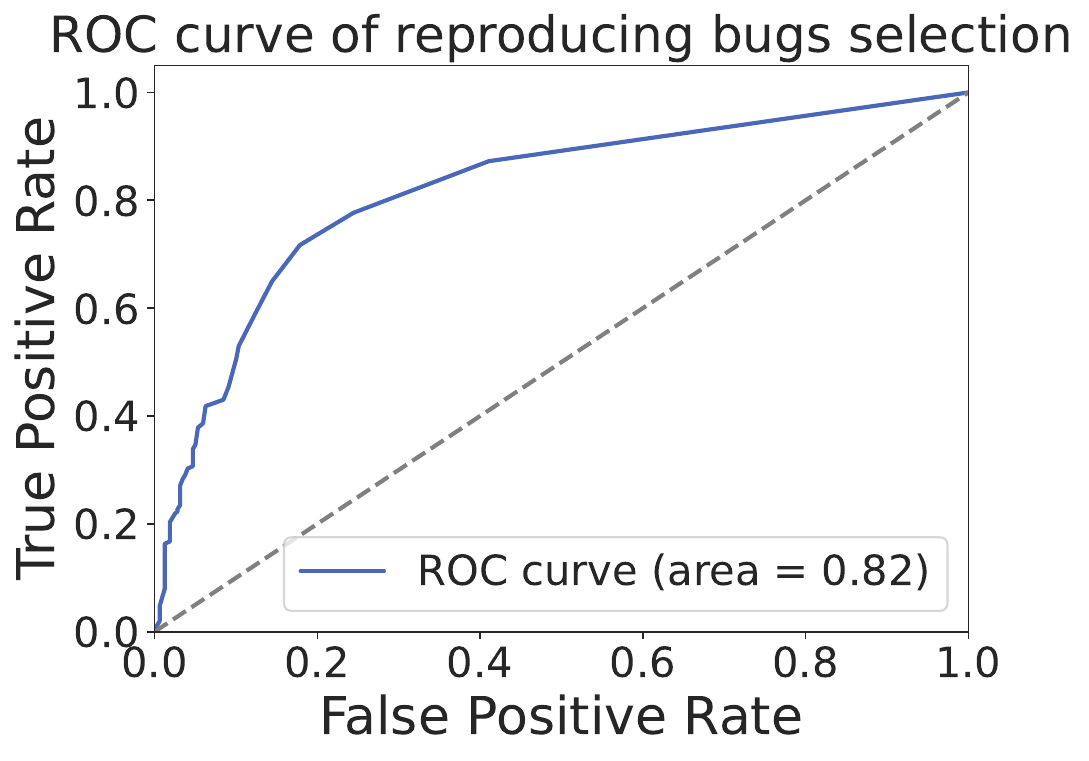}
    \end{subfigure}%
    \begin{subfigure}{0.243\textwidth}
        \centering
        \includegraphics[width=\linewidth]{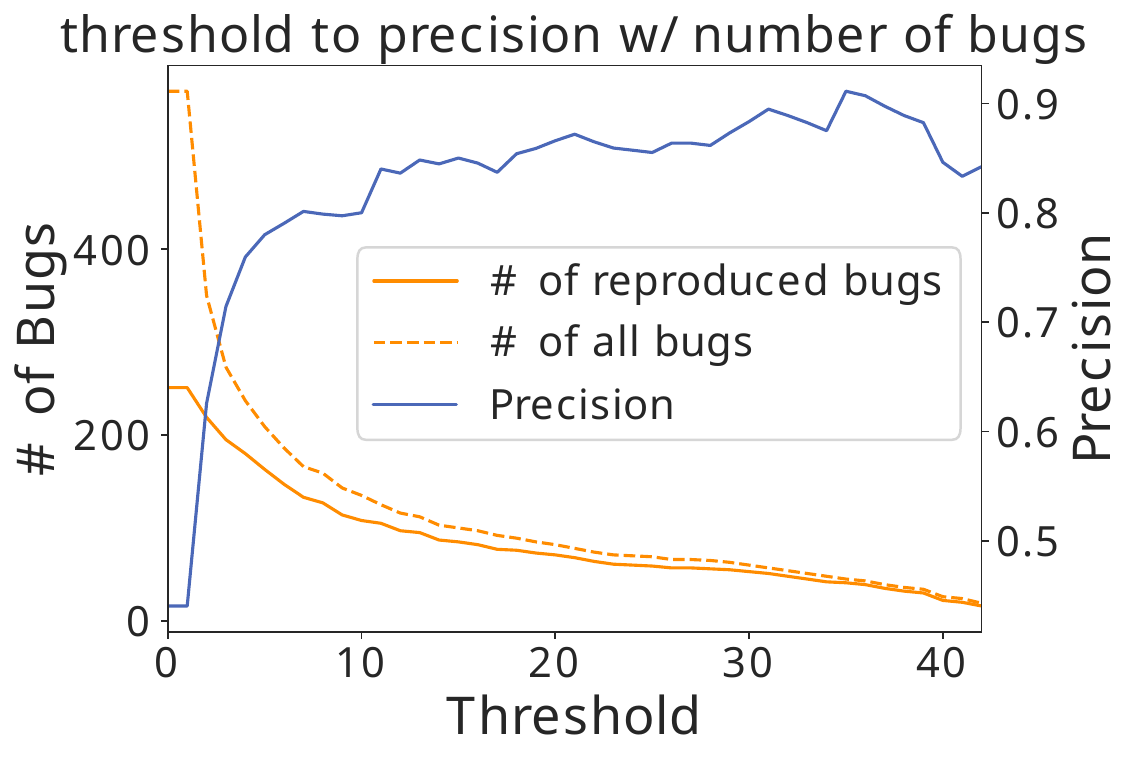}
    \end{subfigure}%
    \caption{ROC curve of bug selection (Left), Effect of thresholds to the number of bugs selected and precision (Right)}
    \label{fig:thr2precision}
\end{figure}

With this research question, we measure how effectively \name prioritizes bug reproducing tests 
via its selection and ranking procedure.
As \name only shows results above a certain agreement threshold, 
$Thr$ from \Cref{sec:sel_n_rank}, we first present the
trade-off between the number of total bugs reproduced and precision 
(i.e., the proportion of successfully reproduced bugs among all selected by \name)
in Figure~\ref{fig:thr2precision}. 
As we increase the threshold, more suggestions (including BRTs) are discarded,
but the precision gets higher, suggesting one can smoothly increase precision 
by tuning the selection threshold. 

We specifically set the agreement threshold to $1$, a conservative value, in order to preserve as many 
reproduced bugs as possible. Among the 570 bugs with a FIB, 350 bugs are selected. Of those 350, 219 are
reproduced (leading to a precision of $0.63 (=\frac{219}{350})$ 
whereas recall (i.e., proportion of selected reproduced bugs among all
reproduced bugs) is $0.87 (=\frac{219}{251})$. 
From the opposite perspective, the selection process 
filters out 188 bugs that were not reproduced, while dropping only a 
few successfully reproduced bugs. Note that if we set the threshold 
to $10$, a more aggressive value, we can achieve a higher precision of $0.84$ 
for a recall of $0.42$. In any case, as \Cref{fig:thr2precision} presents, our 
selection technique is significantly better than random, indicating it can save 
developer resources.

Among the selected bugs, we assess how effective the test rankings of \name are 
over a random baseline. The random approach randomly ranks the syntactic 
clusters (groups of syntactically equivalent FIB tests) of the generated tests.
We run the random baseline 100 times and average the results. 

Table~\ref{tab:ranking_performance} presents the ranking evaluation results. On 
the Defects4J benchmark, the ranking technique of \name improves upon the 
random baseline across all of the $acc@n$ metrics, presenting 30, 14, and 7 more BRTs than the random baseline on $n=1$, $3$, and $5$ respectively.
Regarding $acc@1$, the first column shows 
that 43\% of the top ranked tests produced by \name successfully reproduce the 
original bug report on the first try. When $n$ increases to 5, BRTs can be 
found in 57\% of the selected bugs, or 80\% of all 
reproduced bugs. The conservative threshold choice here,  
emphasizes recall over precision. However, if the threshold is raised, 
the maximum precision can rise to 0.8 (for $Thr=10$, $n=5$).

The $wef@n_{agg}$ values are additionally reported by both summing and 
averaging the $wef@n$ of all (350) selected bugs. The summed $wef@n$ value 
indicates the total number of non-BRTs that would be manually examined within 
the top $n$ ranked tests. Smaller $wef@n$ values indicate that a technique
delivers more bug reproducing tests. Overall, the ranking of 
\name saves up to 14.5\% of wasted effort when compared to the random baseline, 
even after bugs are selected. Based on 
these results, we conclude that \name can reduce wasted inspection effort and 
thus be useful to assist developers.

\begin{table}[t]
    \caption{Ranking Performance Comparison between \name and Random Baseline\label{tab:ranking_performance}}
    \scalebox{0.74}{
    \setlength\tabcolsep{4.2pt}
    \begin{tabular}{@{}l|ll|ll|ll|ll@{}}
    \toprule
    & \multicolumn{4}{c|}{Defects4J}   & \multicolumn{4}{c}{GHRB}                                   \\ \midrule
    & \multicolumn{2}{c|}{$acc@n$ ($precision$)}     & \multicolumn{2}{c|}{$wef@n_{agg}$}     & \multicolumn{2}{c|}{$acc@n$ ($precision$)}     & \multicolumn{2}{c}{$wef@n_{agg}$}     \\ \midrule
    $n$ & \name & random & \name & random & \name & random & \name & random \\ \midrule
    $1$   & \textbf{149} (\textbf{0.43})   & 116 (0.33)  & \textbf{201} (\textbf{0.57})   & 234 (0.67) &  \textbf{6} (\textbf{0.29})  &  4.8 (0.23)   &  \textbf{15} (\textbf{0.71}) &  16.2 (0.77)   \\
    $3$   & \textbf{184} (\textbf{0.53})   & 172 (0.49)  & \textbf{539} (\textbf{1.54})   & 599 (1.71) &  \textbf{7} (\textbf{0.33})  &  6.6 (0.31)   &  \textbf{42} (\textbf{2.0})  &  44.6 (2.12)   \\
    $5$  & \textbf{199} (\textbf{0.57})    & 192 (0.55)  & \textbf{797} (\textbf{2.28})   & 874 (2.5) &   \textbf{8} (\textbf{0.38})  &  7.3 (0.35)  &   \textbf{60} (\textbf{2.86}) &  64.3 (3.06)    \\ \bottomrule
    \end{tabular}}
\end{table}

\begin{tcolorbox}[boxrule=0pt,frame hidden,sharp corners,enhanced,borderline north={1pt}{0pt}{black},borderline south={1pt}{0pt}{black},boxsep=2pt,left=2pt,right=2pt,top=2.5pt,bottom=2pt]
    \textbf{Answer to RQ2-3:} \name can reduce both
    the number of bugs and tests that must be inspected: 33\% of the bugs are safely discarded while preserving 87\% of the successful bug reproduction. Among selected bug sets, 80\% of all bug reproductions can be found within 5 inspections. 
\end{tcolorbox}  
\subsection{RQ3. How well would \name work in practice?}

\begin{table}[ht]
    \centering
    \caption{Bug Reproduction in GHRB:
    x/y means x reproduced out of y bugs\label{tab:dhr_performance}}
    \scalebox{0.94}{
    \begin{tabular}{lr|lr|lr}
    \toprule
    Project & rep/total & Project & rep/total & Project & rep/total\\\midrule
    AssertJ & 3/5 & Jackson & 0/2 & Gson & 4/7 \\
    checkstyle & 0/13 & Jsoup & 2/2 & sslcontext & 1/2 \\
    \bottomrule
    \end{tabular}
    }
\end{table} 

\subsubsection{RQ3-1} We explore the performance
of \name when operating on the GHRB dataset of recent bug reports. 
We find that of the 31 bug reports we study, \name can automatically 
generate bug reproducing tests for 10 bugs based on 50 trials, for a 
success rate of \textbf{32.2\%}. This success rate is similar to the results
from Defects4J presented in RQ1-1, suggesting \name generalizes to new
bug reports. A breakdown of results by project is provided in 
\Cref{tab:dhr_performance}. Bugs are successfully reproduced in AssertJ, Jsoup, 
Gson, and sslcontext, while they were not reproduced in the other two. We could 
not reproduce bugs from the Checkstyle project, despite it having a large 
number of bugs; upon inspection, we find that this is because the project's 
tests rely heavily on external files, which \name has no access to, as shown in 
\Cref{sec:example_output}. \name also does not generate BRTs for the Jackson project,
but the small number of bugs in the Jackson
project make it difficult to draw conclusions from it.

\begin{tcolorbox}[boxrule=0pt,frame hidden,sharp corners,enhanced,borderline north={1pt}{0pt}{black},borderline south={1pt}{0pt}{black},boxsep=2pt,left=2pt,right=2pt,top=2.5pt,bottom=2pt]
    \textbf{Answer to RQ3-1:} \name is capable of generating bug reproducing
    tests even for recent data, suggesting it is not simply remembering 
    what it trained with. 
\end{tcolorbox}  

\subsubsection{RQ3-2} \name uses several predictive factors correlated with successful bug reproduction for selecting bugs and ranking tests. In this research question, we check whether the identified patterns based on the Defects4J dataset continue to hold in the recent GHRB dataset. 

\begin{figure}[ht]
    \centering
    \begin{subfigure}{0.24\textwidth}
        \centering
        \includegraphics[width=\linewidth]{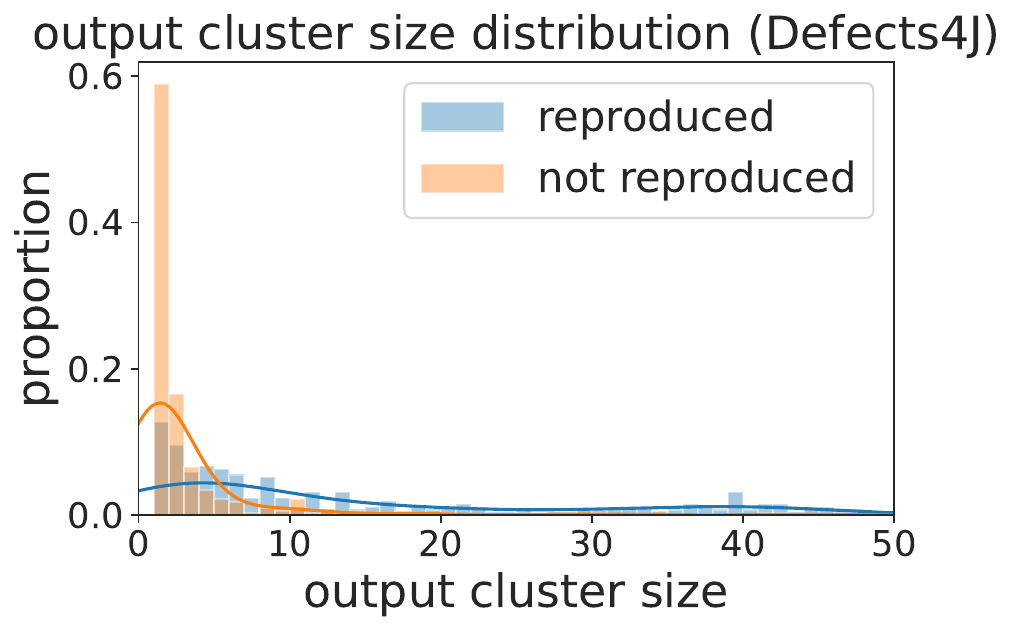}
    \end{subfigure}%
    \begin{subfigure}{0.24\textwidth}
        \centering
        \includegraphics[width=\linewidth]{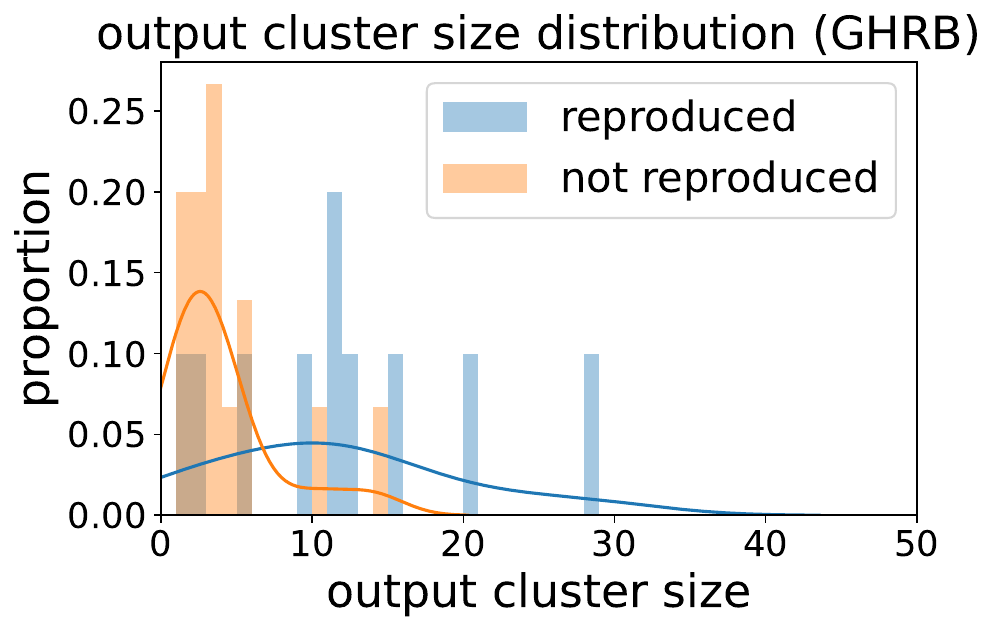}
    \end{subfigure}
    \caption{Distribution of the $max\_output\_clus\_size$ values for reproduced and not-reproduced bugs}
    \label{fig:output_cluster_size_pattern}
\end{figure}

Recall that we use the maximum output cluster size as a measure of agreement among the FIBs, and thus as a selection 
criterion to identify whether a bug has been reproduced. 
To observe whether the criterion is a reliable indicator to 
predict the success of bug reproduction, we observe the trend of $max\_output\_clus\_size$ between the 
two datasets, with and without BRTs. In Figure~\ref{fig:output_cluster_size_pattern}, we see that 
the bugs with no BRT typically have small $max\_output\_clus\_size$, mostly under ten; this pattern is consistent in both datasets.

The ranking results of GHRB are also presented in Table~\ref{tab:ranking_performance}. They are consistent to the results from Defects4J, indicating the features used for our ranking strategy continue to be good indicators of successful bug reproduction.

\begin{tcolorbox}[boxrule=0pt,frame hidden,sharp corners,enhanced,borderline north={1pt}{0pt}{black},borderline south={1pt}{0pt}{black},boxsep=2pt,left=2pt,right=2pt,top=2.5pt,bottom=2pt]
    \textbf{Answer to RQ3-2:} The factors used for the ranking and selection of \name consistently predict bug reproduction in real-world data.
\end{tcolorbox}

\subsection{RQ4. How does the choice of LLM influence \name performance?}

\subsubsection{RQ4-1}

\begin{figure}[ht]
    \centering
    \begin{subfigure}{0.49\textwidth}
        \centering
        \includegraphics[width=\linewidth]{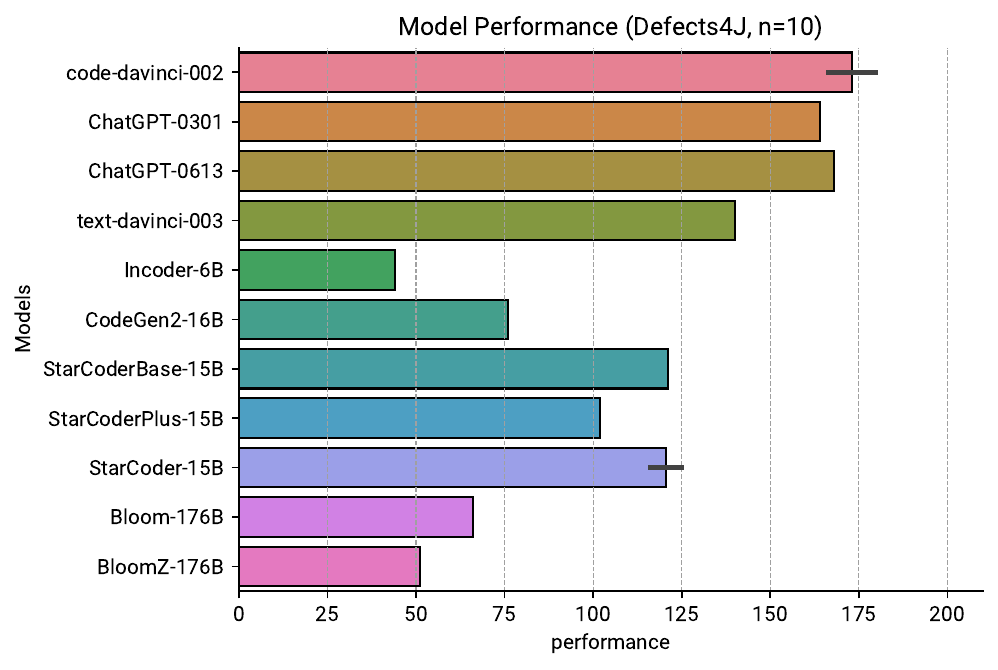}
        \caption{LLM comparison on the Defects4J dataset.}
    \end{subfigure} \\
    \begin{subfigure}{0.49\textwidth}
        \centering
        \includegraphics[width=\linewidth]{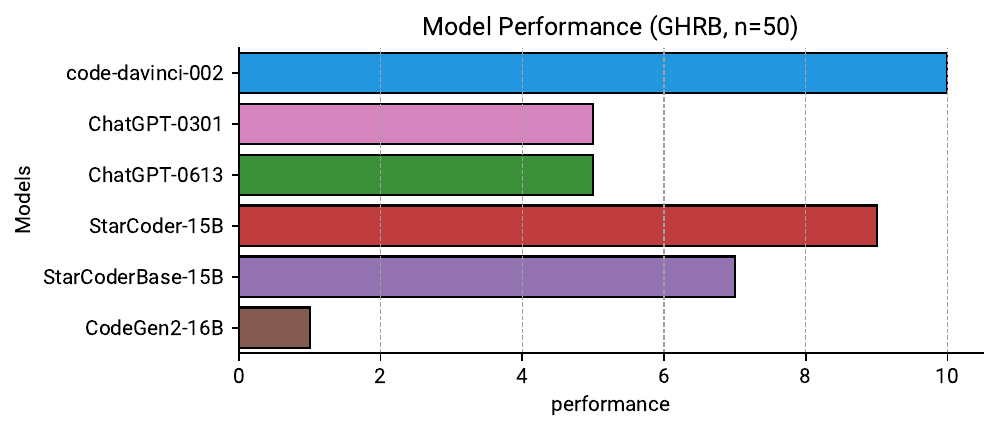}
        \caption{LLM comparison on the GHRB dataset.}
    \end{subfigure}
    \caption{LLM to reproduction performance, with the upper graph depicting Defects4J performance, and the right graph depicting GHRB performance.}
    \label{fig:llm2perf}
\end{figure}

\begin{figure}[ht]
    \centering
    \includegraphics[width=\linewidth]{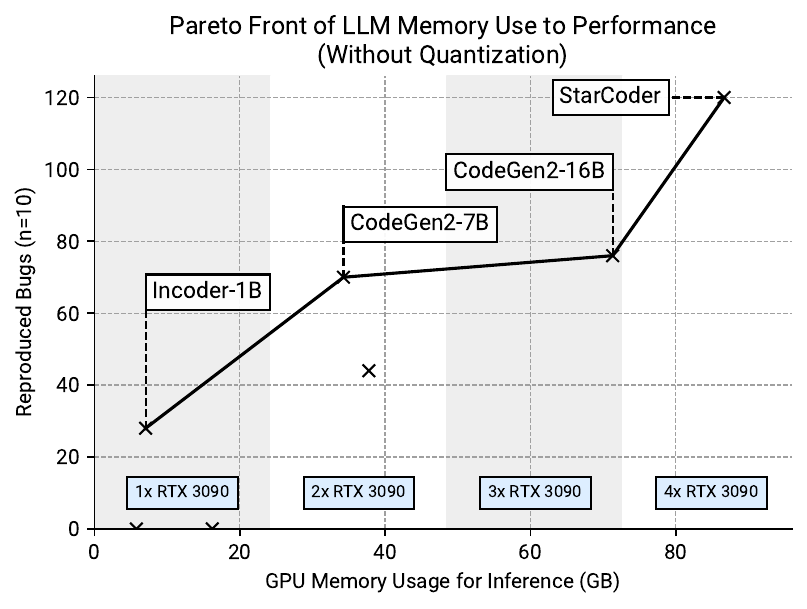}
    \caption{LLM GPU memory use to performance. The graph shows the Pareto front of performance between memory use and performance.}
    \label{fig:perf2memory_pareto}
\end{figure}

While the previous research questions focused on the performance of \name when using one of the first code-based LLMs (Codex), subsequent results evaluate how much \name performance changes when using LLMs of different sizes and training regimes, with a particular focus on open-source LLMs 
that are available to academia. While some work has evaluated the performance of open-source 
LLMs~\cite{jiang2023impact}, since then, multiple open-source LLMs have been introduced thanks to the 
increased interest in LLMs~\cite{li2023starcoder, nijkamp2023codegen2}, which to the best of our knowledge have not yet been evaluated 
in the software engineering literature. 

Figure~\ref{fig:llm2perf} shows the performance of multiple LLMs when given two examples in the prompt. The 
best performing LLM was Codex, which could reproduce a median of 173 bugs given 10 test generation 
attempts for Defects4J dataset. 
Comparing the performance of the open-source models, we find that the StarCoder family of models 
shows the best performance, with the best model, StarCoder-15B, being capable of reproducing 125 
bugs, or 73\% of the performance of Codex. This approaches the results of the currently available OpenAI
models, and could provide stable performance, unlike using the ChatGPT models, which 
may yield different results without notice. Comparing the StarCoder family of models, although 
StarCoder is a fine-tuned version of StarCoderBase trained on Python, this training did not degrade
performance when reproducing Java bugs, indicating that fine-tuning on one language does not 
necessarily hurt performance on other languages. Meanwhile, StarCoderPlus, which is a fine-tuned 
version of StarCoder trained on natural language, showed a substantially worse performance, indicating 
that training on natural language can hurt performance on code-related tasks. This can also be seen in 
the Bloom family of models: the fine-tuned BloomZ model performed substantially worse on 
reproduction than the original Bloom model. Overall, we find that the open-source LLMs can also 
reproduce a substantial number of bugs (albeit lower than the OpenAI models), and thus are a viable 
option when security is a greater concern than performance. As StarCoder showed the 
best performance among the open-source LLMs, we use StarCoder for the subsequent experiments on 
temperature.

By combining this data with our measurement of the memory consumption of each model, we plot
the tradeoff between GPU memory usage and performance in \Cref{fig:perf2memory_pareto}. As expected,
there is a trend of better performance as more GPU memory is used. The four models on the Pareto front
(Incoder-1B, CodeGen2-7B, CodeGen2-16B, StarCoder) are highlighted as well; these are models such that
their performance is nonzero, and there is no model that both performs better and uses less memory.
Conveniently, each model can be mapped to a different number of GPUs as well as \Cref{fig:perf2memory_pareto} shows. 
This information could be helpful to practitioners/researchers when they make
decisions on which LLM to deploy based on their GPU situation.

\begin{tcolorbox}[boxrule=0pt,frame hidden,sharp corners,enhanced,borderline north={1pt}{0pt}{black},borderline south={1pt}{0pt}{black},boxsep=2pt,left=2pt,right=2pt,top=2.5pt,bottom=2pt]
    \textbf{Answer to RQ4-1:} The performance of \name is influenced significantly by the LLM used. Codex shows 
    the best performance of all LLMs, and StarCoder shows the best performance among the open-source 
    LLMs. With less GPUs, Incoder-1B or CodeGen2-7B models are good options as well.
\end{tcolorbox}

\subsubsection{RQ4-2}

To evaluate the performance of \name on the holdout bugs of the GHRB dataset and thus confirm that the LLMs are not simply repeating training data, we select six models that showed strong performance on Defects4J: Codex, GPT-3.5-0301, GPT-3.5-0613, StarCoder, StarCoderBase, and CodeGen2 models. 
As explained in Section~\ref{sec:dataset}, we collected recent bug report and reproducing test pairs after the Codex training data cutoff date, but we find that all the reproducing tests contained in GHRB do not belong to the Stack dataset, which is used to train StarCoder family and CodeGen2 models. We derive this observation from StarCoder's dataset membership test\footnote{\url{https://stack.dataportraits.org/}} provided along with the dataset themselves, a finding also supported by Lee et al.~\cite{lee2023github}.

We generated 50 tests for each bug report with \name for all five models, and the results are presented in Figure~\ref{fig:llm2perf}. The trends of performance observed in Defects4J are observed in GHRB as well, with StarCoder performing the best after code-davinci-002, achieving 90\% of the performance of code-davinci-002 when generating 50 tests. Overall, the average reproduction ratio of StarCoder is 25.2\% for Defects4J and 29.0\% for GHRB in the setting of generating 50 tests, while the GPT-3.5 models reproduced about 22\% of bug reports from Defects4J and 16.1\% of GHRB when generating 50 tests. As LLMs tend to show similar performance for the GHRB data which is likely not part of the training dataset of any LLM, we suggest that \name with general LLMs can be used for novel bug reproduction.

\begin{tcolorbox}[boxrule=0pt,frame hidden,sharp corners,enhanced,borderline north={1pt}{0pt}{black},borderline south={1pt}{0pt}{black},boxsep=2pt,left=2pt,right=2pt,top=2.5pt,bottom=2pt]
    \textbf{Answer to RQ4-2:} LLMs can still perform well for held-out bugs; similarly to Defects4J, StarCoder shows the best performance among the open-source models.
\end{tcolorbox}

\begin{table}[]
    \centering
    \caption{OpenAI model performance under prompts\label{tab:openai_model_changes}}
    \begin{tabular}{lccc}
    \toprule
    Model       & GPT-0301                & GPT-0613               & GPT-0613                \\
    Prompt      & Prompt 1                & Prompt 1               & Prompt 2                \\ \midrule
    Performance & 164 & 72 & 168 \\
    \bottomrule
    \end{tabular}
\end{table}

\subsubsection{RQ4-3}
While the LLMs of OpenAI are the most well-known and show strong 
performance on a multitude of tasks~\cite{brown2020language}, there are few details known about the models, 
particularly starting with the most recent model, GPT-4~\cite{openai2023gpt4}, which did not provide even basic 
details about the model such as model size. Furthermore, OpenAI LLMs are regularly updated, and thus pose 
a challenge for reproducibility in academic research. For example, the LLM that was used in our initial 
experiments, Codex (code-davinci-002), has since become inaccessible to the public. 

Comparing the OpenAI LLM models, gpt-3.5-turbo-0301 achieved a similar 
performance of 164 bugs given 10 test generation attempts, but gpt-3.5-turbo-0613 achieved a much 
worse performance than both of these models, only reproducing 72 bugs under the same condition, as shown
in \Cref{tab:openai_model_changes}.
Initially, such results may appear to represent a shift in model performance, as has been suggested by 
Chen et al.~\cite{chen2023chatgptchange} which noted that the number of executable Python scripts generated by ChatGPT had 
reduced. Inspecting the results from gpt-3.5-turbo-0613, we find that gpt-0613 would generate full test 
files instead of test methods, so that the generated code could no longer be processed correctly by our 
postprocessing pipeline. 
Modifying the prompt by placing the examples in the system message and emphasizing the need to generate
test methods instead of test files, gpt-3.5-turbo-0613 could achieve similar performance to its earlier
version.
Thus, it is difficult to conclude from our data that ChatGPT has become ``worse'' 
over time, as Chen et al.~\cite{chen2023chatgptchange} argue. Rather, as noted by Narayanan and Kappor~\cite{Narayanan2023GPT4worse}, it highlights the risk when building services on top of 
ChatGPT: its behavior can change at any time, and thus postprocessing pipelines or prompts may need to 
adapt without warning.

\begin{tcolorbox}[boxrule=0pt,frame hidden,sharp corners,enhanced,borderline north={1pt}{0pt}{black},borderline south={1pt}{0pt}{black},boxsep=2pt,left=2pt,right=2pt,top=2.5pt,bottom=2pt]
    \textbf{Answer to RQ4-3:} Similarly to prior work, we observe a change in ChatGPT behavior; in our case, ChatGPT became less susceptible to few-shot learning, and our post-processing pipeline which relied on a specific output format failed.
\end{tcolorbox}

\subsubsection{RQ4-4}
\label{sec:cgpt_change_results}

\begin{figure}[ht]
    \centering
    \begin{subfigure}{0.24\textwidth}
        \centering
        \includegraphics[width=\linewidth]{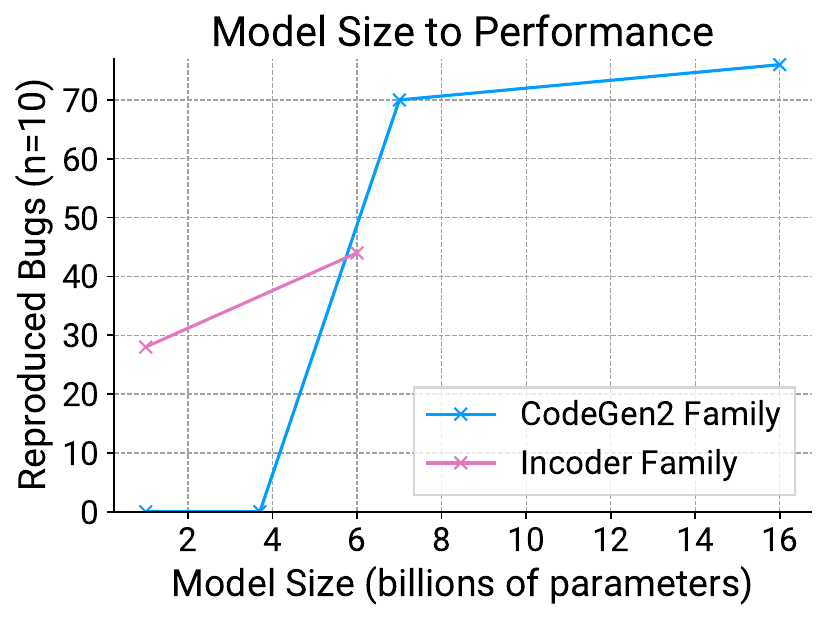}
        \caption{Model Size}
        \label{fig:llmsize2perf}
    \end{subfigure}%
    \begin{subfigure}{0.24\textwidth}
        \centering
        \includegraphics[width=\linewidth]{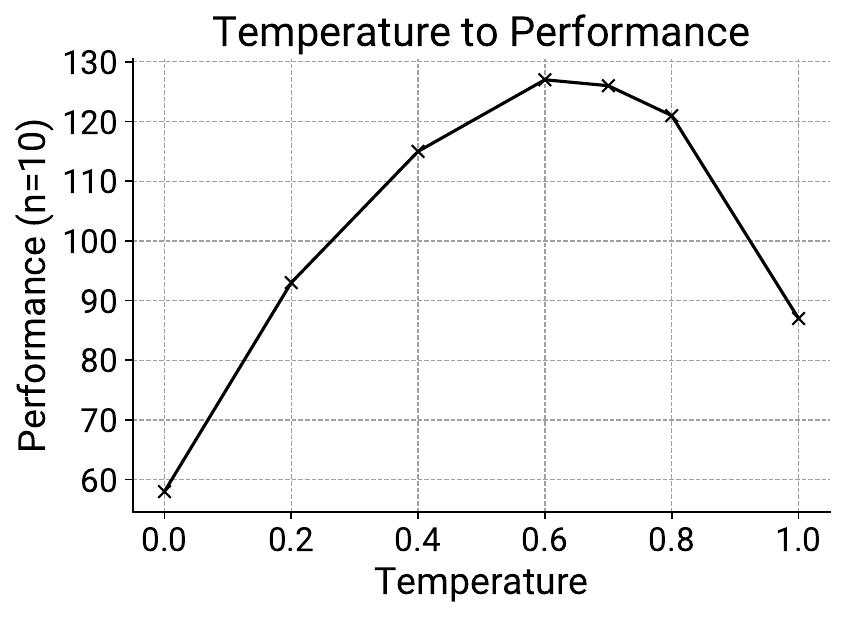}
        \caption{Temperature}
        \label{fig:samptemp2perf}
    \end{subfigure}
    \caption{Evaluation of the influence of LLM configuration to performance.}
    \label{fig:selection_thr}
\end{figure}

While \Cref{fig:llm2perf} compared the performance of LLMs trained in different ways, we also make a comparison between LLMs that are from the same family and were thus trained in a similar manner, but are of substantially different size, to demonstrate how LLM size can affect bug reproduction performance. We plot the results of these experiments in \Cref{fig:llmsize2perf}. As the graph shows, bug reproduction suddenly becomes possible when using 
the 7B model for CodeGen2. Such results are reminiscent of `emergent' properties of LLMs~\cite{wei2022emergent},
in which LLM capabilities suddenly appear at a certain model size, which makes LLM capabilities difficult
to predict prior to training.
On the other hand, in the Incoder family, even the 1B model can reproduce a certain amount of bugs
using our default prompt.
Regardless of whether the property is emergent, the results in \Cref{fig:llm2perf} show that 
bug reproduction performance tends to increase as model size increases.

\begin{tcolorbox}[boxrule=0pt,frame hidden,sharp corners,enhanced,borderline north={1pt}{0pt}{black},borderline south={1pt}{0pt}{black},boxsep=2pt,left=2pt,right=2pt,top=2.5pt,bottom=2pt]
\textbf{Answer to RQ4-4:} \name performance improves as the underlying LLM size increases; for CodeGen2, a sudden appearance of reproduction capability is observed.
\end{tcolorbox}  

\subsubsection{RQ4-5}
\label{sec:temp_perf_results}

\Cref{fig:samptemp2perf} shows the performance of \name when using StarCoder. As the graph shows, we find 
that the performance was best when the temperature was 0.6, which was similar to our initial setting of 
temperature=0.7. At temperature=0.6, \name-StarCoder could reproduce 127 bugs when 
generating ten tests for each bug report. Looking at each temperature, we find that at temperatures 
lower than 0.6, the LLM tends to generate identical or similar tests for a given bug report, and thus does 
not reproduce more bugs as more tests are generated. Meanwhile, for higher temperatures, the
coherence of the LLM-generated results deteriorates, and thus increasingly less bugs are reproduced. 
Indeed, while not shown in the graph, our experiments when the temperature was set to 2.0 revealed 
that the LLM would almost exclusively generate unparsable code, indicating that 
setting the LLM to the right temperature is important when achieving strong bug reproduction 
performance. 

\begin{tcolorbox}[boxrule=0pt,frame hidden,sharp corners,enhanced,borderline north={1pt}{0pt}{black},borderline south={1pt}{0pt}{black},boxsep=2pt,left=2pt,right=2pt,top=2.5pt,bottom=2pt]
    \textbf{Answer to RQ4-5:} The performance of \name-StarCoder is optimized when the temperature is set to 0.6, 
    which gets a good balance between generation diversity and result coherence.
\end{tcolorbox}  

\subsection{RQ5. How does the choice of LLM influence the selection and ranking aspect of \name?}

\subsubsection{RQ5-1}

\begin{figure}[ht]
    \centering
    \begin{subfigure}{0.24\textwidth}
        \centering
        \includegraphics[width=\linewidth]{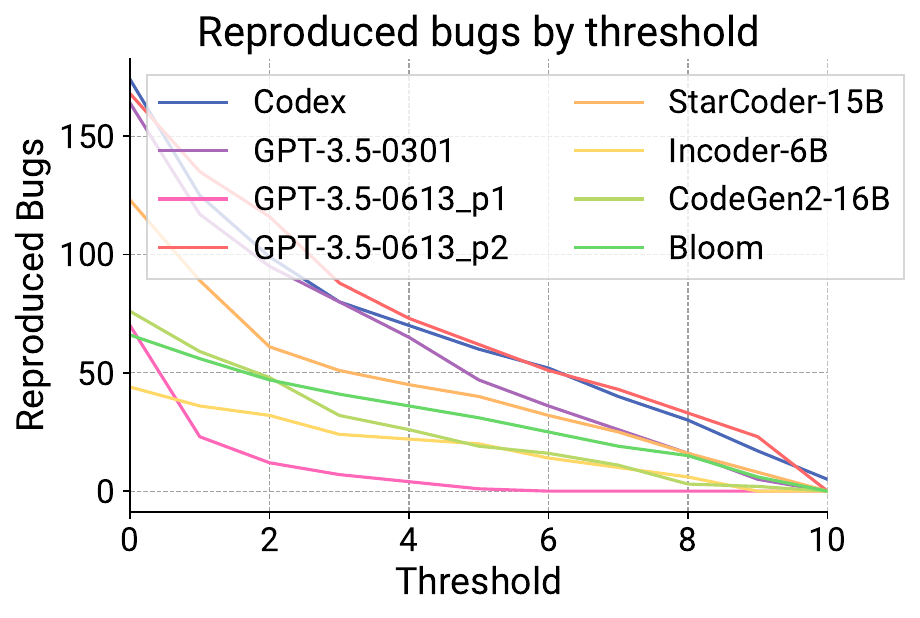}
    \end{subfigure}%
    \begin{subfigure}{0.24\textwidth}
        \centering
        \includegraphics[width=\linewidth]{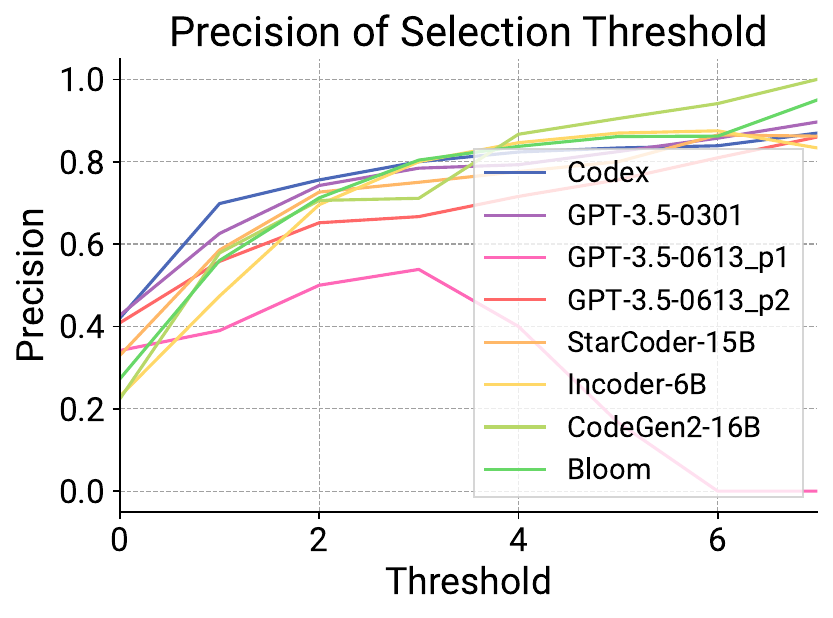}
    \end{subfigure}
    \caption{Number of reproduced bugs and precision by selection with varying thresholds.}
    \label{fig:selection_thr}
\end{figure}

As \name selects bugs and prioritizes tests to reduce the number of bugs and tests that need to be inspected, we further investigate how different LLMs perform when using the selection and ranking strategy. We selected the best-performing models from each model family and focused our evaluation
on those models for simplicity.

Figure~\ref{fig:selection_thr} shows that the number of reproduced bugs and precision (i.e., the number of reproduced bugs among the selected ones divided by the number of selected bugs) with varying thresholds. As the threshold rises, the number of reproduced bugs decrease, while precision generally increases (with the exception of gpt-3.5-turbo-0613).
Based on the result, we suggest that \name's selection criterion, maximum size of output clusters, is a useful indicator for successful bug reproduction regardless of the LLM used. In addition, the abnormally low performance
of gpt-3.5-turbo-0613 when using Prompt 1 (\Cref{fig:selection_thr}, GPT-3.5-0613\_p1; for discussion, see \Cref{sec:cgpt_change_results}) is accompanied by abnormal
threshold-precision behavior, in which increasing the selection threshold does not lead to improved
precision. Such abnormalities may potentially used as diagnostics for LLM-based applications, which
may provide insight that raw performance may not provide.

\begin{table}[ht]
\scalebox{0.9}{
    \begin{tabular}{l|rrr|rrr}
        \toprule
                  model &  acc@1 &  acc@3 &  acc@5 &  prec@1 &  prec@3 &  prec@5 \\
        \midrule
                  Codex &    107 &    122 &    123 &    0.60 &    0.68 &    0.69 \\
        GPT-3.5-0613\_p2 &    106 &    125 &    134 &    0.44 &    0.52 &    0.55 \\
           GPT-3.5-0301 &     92 &    112 &    115 &    0.49 &    0.60 &    0.61 \\
          StarCoder-15B &     70 &     83 &     89 &    0.46 &    0.55 &    0.59 \\
           CodeGen2-16B &     48 &     56 &     57 &    0.47 &    0.55 &    0.56 \\
                  Bloom &     45 &     54 &     56 &    0.45 &    0.54 &    0.56 \\
             Incoder-6B &     29 &     33 &     34 &    0.38 &    0.43 &    0.45 \\
        GPT-3.5-0613\_p1 &     18 &     22 &     23 &    0.31 &    0.37 &    0.39 \\
        \bottomrule
        \end{tabular}}
\caption{Ranking performance of \name with different sampling temperatures (selection threshold $= 1$, $n=10$)}
\label{tab:llms_ranking_performance_d4j}
\end{table}

Table~\ref{tab:llms_ranking_performance_d4j} shows the performance of \name after ranking FIBs among the selected bugs with a threshold of 1. Codex is still the best performing LLM after ranking, followed by the ChatGPT and StarCoder models. The performance of CodeGen2 and Bloom models are similar, and Incoder-6B shows the worst performance. The best performing open-source LLM, StarCoder, achieves about 60\% precision when inspecting top five ranked tests. Overall, the relative performance of LLMs is preserved after additional postprocessing steps, and consistently improves the precision for all LLMs.

\begin{tcolorbox}[boxrule=0pt,frame hidden,sharp corners,enhanced,borderline north={1pt}{0pt}{black},borderline south={1pt}{0pt}{black},boxsep=2pt,left=2pt,right=2pt,top=2.5pt,bottom=2pt]
    \textbf{Answer to RQ5-1:} The selection and ranking strategy of \name is robust to the choice of LLM.
\end{tcolorbox}

\subsubsection{RQ5-2}
\label{sec:temp_ranking_results}

\begin{figure}[ht]
    \centering
    \begin{subfigure}{0.24\textwidth}
        \centering
        \includegraphics[width=\linewidth]{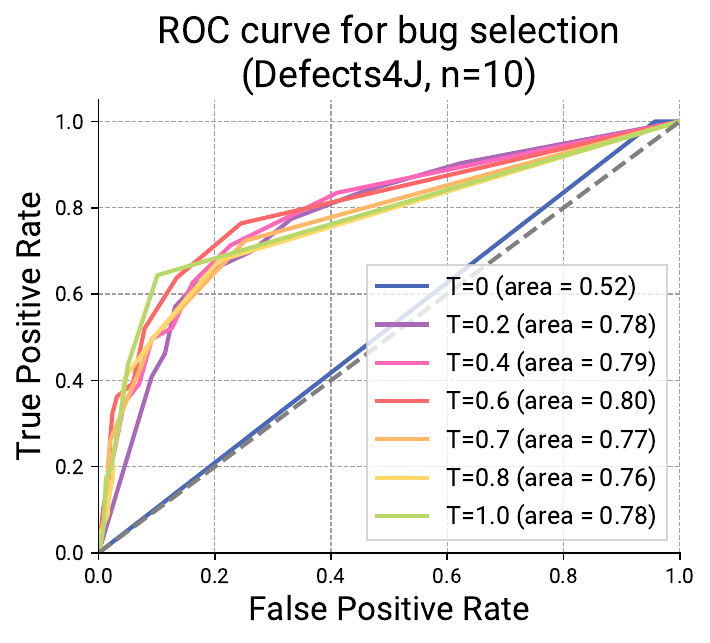}
        \caption{}
    \end{subfigure}%
    \begin{subfigure}{0.24\textwidth}
        \centering
        \includegraphics[width=\linewidth]{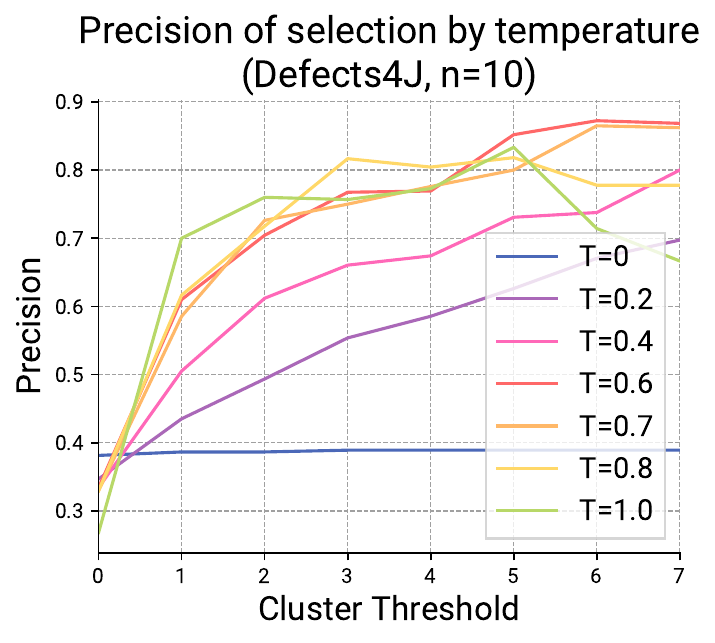}
        \caption{}
    \end{subfigure}
    \caption{Selection performance by temperature.}
    \label{fig:temperature2selection}
\end{figure}

As discussed earlier, the selection of temperature may have a significant effect on the selection
and ranking algorithm of \name, as it directly controls how repetitive multiple runs of \name
will be. Thus, we investigated how selection and ranking performance is affected by the choice
of LLM sampling temperature. Results are shown in \Cref{fig:temperature2selection}. To our
surprise, the selection performance of \name, evaluated via the ROC-AUC metric that
is independent to the selection of threshold, is remarkably consistent for different temperature
values, with all showing an ROC-AUC of 0.76-0.80. We are unaware of prior work reporting
similar results, and are similarly perplexed as to why ROC-AUC should stay so consistent under
different temperature settings. It is clear that different temperature settings yield different
behavior, in the expected way: looking at \Cref{fig:temperature2selection}(b), we see that for lower
temperatures, cluster thresholds need to be larger to have high predictive power, as results
are already very similar due to the low randomness of sampling; meanwhile, for higher temperatures,
small thresholds may already be predictive of high precision as the outputs are diverse, and
the fact that they overlap at all is significant. Nonetheless, these effects are counterbalanced
and all temperatures end up leading to a similar bug selection performance.
The one exception is T=0, as almost all results are identical for different runs in this scenario. (T=0 did not complete test output overlap in four cases; in one case, a test appeared to be
flaky, while in three other cases, the tests would print timestamps, which was not abstracted during
the clustering process.)

\begin{table}[ht]
\centering
\scalebox{0.9}{
    \begin{tabular}{l|rrr|rrr}
        \toprule
        Temperature &  acc@1 &  acc@3 &  acc@5 &  prec@1 &  prec@3 &  prec@5 \\
        \midrule
        T=0.0 &     58 &     58 &     58 &    0.38 &    0.38 &    0.38 \\
        T=0.2 &     78 &     91 &     93 &    0.29 &    0.34 &    0.35 \\
        T=0.4 &     88 &    111 &    115 &    0.26 &    0.32 &    0.33 \\
        T=0.6 &     99 &    122 &    125 &    0.26 &    0.32 &    0.33 \\
        T=0.7 &     97 &    117 &    123 &    0.26 &    0.31 &    0.33 \\
        T=0.8 &     94 &    115 &    120 &    0.26 &    0.31 &    0.33 \\
        T=1.0 &     60 &     85 &     86 &    0.19 &    0.26 &    0.27 \\
        \bottomrule
        \end{tabular}}
\caption{Ranking performance of \name-StarCoder with different sampling temperature (selection threshold $=1$, $n=10$)}
\label{tab:temperature_to_ranking}
\end{table}

Ranking results by temperature are shown in \Cref{tab:temperature_to_ranking}. Similarly
to the overall reproduction results presented in \Cref{fig:samptemp2perf}, sampling at T=0.6 shows
the best performance, even considering both selection and ranking. Meanwhile, as apparent on the
right side of \Cref{tab:temperature_to_ranking}, the precision of \name decreases as the temperature
increases, which may be related to the tendency for lower temperature outputs to generally yield better performance on a single run~\cite{wang2023selfconsistency}.

\begin{tcolorbox}[boxrule=0pt,frame hidden,sharp corners,enhanced,borderline north={1pt}{0pt}{black},borderline south={1pt}{0pt}{black},boxsep=2pt,left=2pt,right=2pt,top=2.5pt,bottom=2pt]
    \textbf{Answer to RQ5-2:} Surprisingly, while results sampled from different temperatures have different characteristics, the performance of our selection algorithm is similar regardless of temperature. Considering all post-processing, T=0.6 showed the best performance, similarly to RQ4-4.
\end{tcolorbox}

\section{Discussion}
\label{sec:discussion}

\subsection{Example output of \name} 
\label{sec:example_output}
Example outputs of \name-Codex from the GHRB dataset are presented to provide further context for our quantitative results.

\begin{table}[h!]
    \caption{\label{tab:assertj-success} Bug Report Successfully Reproduced: URLs are omitted for brevity (AssertJ-Core Issue \#2666)}
    \begin{tabular}{@{}lp{0.4\textwidth}@{}}
    \toprule
    \textbf{Title}     & \textbf{assertContainsIgnoringCase fails to compare i and I in tr\_TR locale} \\ \midrule
    \multicolumn{2}{l}{\begin{tabular}[c]{@{}l@{}} See org.assertj.core.internal.Strings\#assertContainsIgnoringCase \\

        [url] \\

        I would suggest adding [url] verification to just ban \\
        toLowerCase(), toUpperCase() and other unsafe methods: \#2664\end{tabular}} \\ \bottomrule
    \end{tabular}
\end{table}

\begin{lstlisting}[basicstyle=\footnotesize\ttfamily,
    float=t,
    columns=flexible,
    breaklines=true,
    language=java,
    caption={Generated FIB test for AssertJ-Core-2666.},
    label={lst:assertj-succ-test},]
public void testIssue952() {
  Locale locale = new Locale("tr", "TR");
  Locale.setDefault(locale);
  assertThat("I").as("Checking in tr_TR locale").containsIgnoringCase("i");
}
\end{lstlisting}

\Cref{tab:assertj-success} shows our first example, a successful reproduction
of issue \#2685 from the AssertJ-Core project. The issue reports a bug for a
particular locale (\code{tr\_TR}), where the letter \code{I} is being mishandled.
A successful test reproducing this issue from \name is presented in Listing~\ref{lst:assertj-succ-test};
it is noteworthy that \name successfully generated this test even though the bug
report does not include any executable code. Instead, it used the method names
(\code{containsIgnoringCase}) provided within the bug report to generate a useful test.
Furthermore, for this bug, a BRT is ranked at second place, meaning that a developer
could quickly discover and use it.

\begin{table}[h!]
    \caption{\label{tab:checkstyle-failure} Bug Report Reproduction Failure: Lightly edited for clarity (Checkstyle Issue \#11365)}
    \begin{tabular}{@{}lp{0.4\textwidth}@{}}
    \toprule
    \textbf{Title}     & \textbf{FinalClassCheck: False positive with anonymous classes} \\ \midrule
    \multicolumn{2}{l}{\begin{tabular}[c]{@{}l@{}} ... I have executed the cli and showed it below, as cli describes the \\
        problem better than 1,000 words\\
        $\rightarrow$\code{src cat Test.java} \\
        \code{[...]} \\
        \code{public class Test \{} \\
        \hspace*{1em}\code{class a \{ // expected no violation} \\
        \hspace*{2em}\code{private a() \{\} \} \}} \\
        \code{[...]} \\
        $\rightarrow$\code{java [...] -c config.xml Test.java} \\
        \code{Starting audit...} \\
        \code{[ERROR] Test.java:3:5: Class a should be declared as final.} \\
        \end{tabular}} \\ \bottomrule
    \end{tabular}
\end{table}

\begin{figure}
\begin{lstlisting}[basicstyle=\footnotesize\ttfamily,
    columns=flexible,
    breaklines=true,
    language=java,
    caption={Generated FIB test for Checkstyle-11365.},
    label={lst:checkstyle-fail-test},]
public void testFinalClass() throws Exception {
    final DefaultConfiguration checkConfig =
            createModuleConfig(FinalClassCheck.class);
    final String[] expected = CommonUtil.EMPTY_STRING_ARRAY;
    verify(checkConfig, getPath("InputFinalClassAnonymousClass.java"), expected);
}
\end{lstlisting}
\end{figure}

Our second example, for which the bug report is presented in \Cref{tab:checkstyle-failure}, 
is a case in which \name failed to successfully reproduce the bug. Here, Checkstyle makes
the incorrect inference that a particular class should be declared \code{final}, which leads
to misleading results. One of the FIB tests generated by \name for this bug report is
presented in Listing~\ref{lst:checkstyle-fail-test}. Because the Java file the test
refers to on Line 5 does not exist within the project, the test fails for an inaccurate
reason. Thus, this test makes a good cautionary example to illustrate the limitations of
\name, as it cannot change the execution environment of tests, which may be a prerequisite
to reproduce a bug. Nonetheless, the test itself is functional; by inserting the Java code provided within
the bug report into the appropriate file and re-running the test, the test is capable of
reproducing the bug. Such results suggest that future work may also work on constructing
the execution environment of tests to further widen the amount of bugs that can be automatically
reduced, and thus alleviate developer effort on such tasks.

\subsection{Code Overlap with Bug Report}
\begin{figure}[ht]
    \centering
        \includegraphics[width=0.95\linewidth]{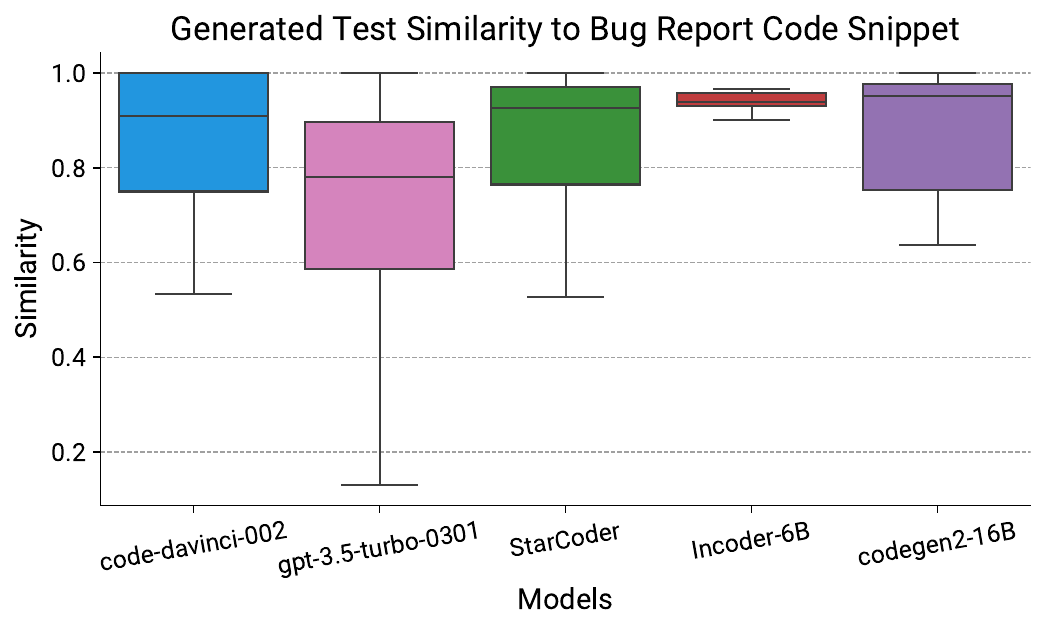}
    \caption{Similarity between generated tests and code snippets contained in bug reports among multiple LLMs.}
    \label{fig:similarity_code_snippet}
\end{figure}

As Just et al. point out~\cite{just2018comparing}, bug reports can already contain partially or fully executable test code, but developers rarely adopt the provided tests as is.
To investigate whether the performance of \name is due to 
efficient extraction of report content or effective
synthesis of test code,
we analyzed the 750 bug reports from Defects4J used in our experiment, and compared them to tests
generated by Codex (code-davinci-002). We find that 19.3\% of them 
had full code snippets (i.e., code parsable to a class or method), while 39.2\% had partial code 
snippets (i.e., not a complete class or method but in the form of source code statements or 
expressions); finally 41.5\% did not contain code snippets inside. 
Considering only the 251 bug reports that 
\name successfully reproduced, the portion of containing the
full snippets got slightly higher (25.1\%), whereas the portion of bug reports with partial
snippets was 37.9\%, and 37.1\% did not have code snippets. When \name generated tests from
bug reports containing any code snippets, we find that on average 81\% of the tokens in the body of
the \name-generated test methods overlapped with the tokens in the code snippets.

In \Cref{fig:similarity_code_snippet}, we evaluate how similar the output of LLMs evaluated were to code snippets provided within the bug report. code-davinci-002, StarCoder, and
CodeGen2 all showed a similar similarity between the generated test and snippets provided in the
bug report, suggesting that they rely on the bug report to a similar degree; judging from the
similarity distribution, these models could successfully reproduce bugs even when the exact reproducing
snippet is not provided within the bug report. Meanwhile, Incoder-6B shows a very tight
distribution around a similarity of 0.9, suggesting that most of its successes were due to very
similar code already being provided within the bug report. Finally, gpt-3.5-turbo-0301 succeeded
even when the similarity to the bug report was low; this may be due to its different training
focus (natural language) and consequently even when the generated snippets did not resemble
the provided code, it could succeed nonetheless.

Although full code snippets are often provided in bug reports, this does not necessarily mean
that such code will successfully reproduce bugs; indeed, the Copy\&Paste baseline succeeded
far less often than \name, as shown in \Cref{fig:baseline-venn}, only reproducing 36 bugs.
Thus, we conclude that while \name can be influenced by how much code is provided within a bug report,
it nonetheless also succeeds in correcting non-reproducing code or even from generating test
code from bug reports almost exclusively in natural language. In turn, this means that the
LLMs used in our study can both extract the helpful parts of code snippets provided in bug reports,
as well as synthesize tests from scratch, based on the given report.

\section{Related Work}
\label{sec:relwork}

\subsection{Test Generation}
Since at least the 1970s, automatic test generation has been a topic of research
in software engineering~\cite{Miller:1976ht}. In the long period since, there have
been multiple shifts in approaches, such as the shift that occurred when
object-oriented programming languages gained popularity, which necessitated
the derivation of method call sequences~\cite{Pacheco:2007oq,
Fraser:2013vn}. An often-cited problem with test generation techniques is the
oracle problem~\cite{Barr:2015qd}, which stipulates the difficulty of
automatically determining what the correct behavior of a software system should be.
As a result of this difficulty, test generation techniques will rely on
so-called implicit oracles such as crashes~\cite{Pacheco:2007oq}, 
or assume a regression testing scenario and treat the current output of the system
as correct~\cite{Fraser:2013vn,Tufano2020ji}. Still others will use rule-based heuristics
to derive oracle behavior from structured specification documents~\cite{Motwani2019Swami};
these techniques may suffer in performance when the structure changes, whereas \name
makes no assumptions about the structure of bug reports.

Meanwhile, there is a body of literature that focuses on reproducing bugs, similarly
to \name. The oracle problem is nonetheless an obstacle, so program crashes are often
used as a proxy for reproducing a bug~\cite{Barr:2015qd}. Most existing work on
reproducing crashes focuses in particular on reproducing a crash stack trace~\cite{Nayrolles2015Jcharming, 
chen2014star, xuan2015crash, Soltani2018aa, Derakhshanfar2020wt}. 
On the other hand, Yakusu~\cite{Fazzini2018aa} and ReCDroid~\cite{Zhao2019na} analyze
formatted bug reports to automatically reconstruct a sequence of actions that leads
to crashes in mobile applications, and encapsulate those sequences in tests.
All the aforementioned work differs from \name, as they use the implicit crash oracle
to discern whether a bug was reproduced and thus can only deal with crash bugs.
Meanwhile, BEE~\cite{Song2020Bee} proposes a technique to parse bug reports and extract
the observed or expected behavior from natural language descriptions, but does
not generate tests.
To the best of our knowledge, we are the first to propose a technique to 
reproduce general bug reports in Java projects.

Since our previous publication~\cite{Kang2023LIBRO}, significant research progress has been
made in generating tests using LLMs. Lemieux et al.~\cite{lemieux2023codamosa} combine LLM
generation into an SBST loop to achieve better results than just using traditional
SBST or LLMs alone. Liu et al.~\cite{Liu2023qtypist} generate text input for tests of mobile applications,
and demonstrate that generating text input can help improve coverage. Closely
related to our work, Feng et al.~\cite{feng2023prompting} propose AdbGPT, a technique that
automatically reproduces bug reports for mobile applications. Our work is
distinct from the aforementioned techniques, as (i) it targets the reproduction of
bugs in general software, and (ii) we provide a comparison of the performance of
a large number of LLMs in generating tests. In particular, we are unaware of
such a comparison of LLMs being made, and thus believe our experiments
may contribute to the software engineering community.

\subsection{Code Synthesis}
Similarly to test generation, code synthesis has been researched for a
long time. The traditional approach to code synthesis was to use SMT
solvers within the framework of Syntax-Guided 
Synthesis (SyGuS)~\cite{Alur2015SyGuS}. Recent research on code synthesis
has also used machine learning techniques, which also yield strong
performance; for example, Chen et al.~\cite{Chen2021ec} demonstrated that
LLMs could effectively synthesize Python code from a natural language
description. To further improve the performance of code synthesis, some
techniques have also generated tests: AlphaCode used automatically 
generated tests to boost their code synthesis performance~\cite{Li2022kp}, 
while CodeT jointly generated tests and code from a natural language 
description~\cite{Chen2022iw}. However, these techniques focus on
code synthesis, not test generation, and thus the tests are discarded
after evaluating the generated code. In contrast, \name is fully focused
on test generation, and furthermore introduces a novel pipeline to
reduce developer effort in inspecting generated results.

\section{Conclusion}
\label{sec:conclusion}

In this paper, we introduce \name, a 
technique that uses a pretrained LLM to analyze bug reports, generate
prospective tests, and finally rank and suggest the generated solutions based 
on a number of simple statistics. Upon extensive analysis, we find that \name using
OpenAI's code-davinci-002 LLM
is capable of reproducing a significant number of bugs in the Defects4J 
benchmark as well as generalize to a novel bug dataset that was not part of its
training data; furthermore, we demonstrated that \name could indicate when its
tests were likely to actually reproduce the bug. Our additional large-scale experiments
comparing the bug reproducing performance of 15 LLMs reveal that open-source
LLMs can also show strong performance, with the StarCoder LLM showing the best
performance among open-source LLMs evaluated, and other confirmations such that
the size of the LLM positively influences bug reproduction performance. Our
evaluation of our selection and ranking techniques also show that they are
capturing general properties of LLMs for bug reproduction, as the heuristics
work in the same manner over all LLMs evaluated. 
We hope that our experiments and results are of use
to both researchers and practitioners when deciding which LLM would be appropriate
for their application, and plan to continue researching the productive
capabilities of open-source LLMs.

\bibliographystyle{IEEEtran}
\bibliography{newref,reference}
\end{document}